\documentclass[12pt]{article}
\usepackage{epsfig, amssymb}
\newcommand{\be}{\begin{equation}}
\newcommand{\ee}{\end{equation}}
\newcommand{\bea}{\begin{eqnarray}}
\newcommand{\eea}{\end{eqnarray}}
\newcommand{\bA}{\begin{array}}
\newcommand{\eA}{\end{array}}
\newcommand{\bc}{\begin{center}}
\newcommand{\ec}{\end{center}}
\newcommand{\al}{\alpha}

\newcommand{\ra}{\rightarrow}

\newcommand{\ie}{{\it i.e.}}
\newcommand{\eg}{{\it e.g.}}

\newcommand{\Rea}{\mathop{\rm Re}}

\newcommand{\Nt}{${\cal N}{=}2$}

\def\BC{{\mathbb C}}
\def\BP{{\mathbb P}}

\def\BZ{{\mathbb Z}}
\def\BQ{{\mathbb Q}}
\def\BM{\mbox{\boldmath$M$}}
\def\BN{\mbox{\boldmath$N$}}

\begin{document}

\begin{titlepage}

\bc

\hfill  {Duke-CGTP-04-09} \\
\hfill  {TIFR-TH-04-26} \\
\hfill  {\tt hep-th/0412337} \\
         [22mm]

{\Huge On tachyons, gauged linear sigma models, \\[6pt] and 
flip transitions}
\vspace{10mm}

{\large David R.~Morrison$^a$ and K.~Narayan$^{a, b}$} \\
\vspace{3mm}
{$^a$\small \it Center for Geometry and Theoretical Physics, \\}
{\small \it Duke University, \\}
{\small \it Durham, NC 27708, USA.\\}
\vspace{1mm}
{$^b$\small \it Tata Institute of Fundamental Research, \\}
{\small \it Homi Bhabha Road, \\}
{\small \it Colaba, Mumbai - 400005, India.\\}
{\small Email: \ drm, narayan@cgtp.duke.edu, \ 
narayan@theory.tifr.res.in}\\

\ec
\medskip
\vspace{25mm}

\begin{abstract}
We study systems of multiple localized closed string tachyons and the
phenomena associated with their condensation, in $\BC^3/\BZ_N$
nonsupersymmetric noncompact orbifold singularities using gauged
linear sigma model constructions, following hep-th/0406039. Our
study reveals close connections between the combinatorics of
nonsupersymmetric flip transitions (between topologically
distinct resolutions of the original singularity), the physics of
tachyons of different degrees of relevance and the singularity
structure of the corresponding residual endpoint geometries. This in
turn can be used to study the stability of the phases of 
gauged linear sigma models and
gain qualitative insight into the closed string tachyon potential.
\end{abstract}

\end{titlepage}

\newpage \begin{tableofcontents}
\end{tableofcontents}

\vspace{10mm}

Both from the point of view of gaining a deeper understanding of the
vacuum structure of string theory as well as other angles such as
understanding the role of time in string theory, the analyses of
tachyon dynamics that have emerged over the last few years following
Sen \cite{sen} have been particularly insightful (see, \eg,
\cite{sen, emilrev, minwalla0405} for general reviews of tachyon
condensation). Ideally one would like to gain insight into the
dynamics of tachyonic instabilities in string theory by studying their
evolution in time: this is however quite hard in general.
Thus one needs to resort to other techniques that are more amenable to
precise calculation. In particular, considerable qualitative insight
into the physics is gained by replacing the on-shell question of time
evolution of a stringy system with the off-shell question of studying
the ``space of string theories'' via off-shell formulations, such as
string field theory (see, \eg, \cite{sen, watiSFT}) or string
worldsheet renormalization group flow approximated by an appropriate
gauged linear sigma model (GLSM) (see, \eg, \cite{wittenIAS,
horiICTP}) and their mirror descriptions (see, \eg, \cite{horiICTP}).

These issues are complicated still further for closed string tachyons
where one could {\it a priori}\/ suspect delocalized tachyonic
instabilities to lead to the decay of spacetime geometry itself, by
analogy with open string tachyons which lead to D-brane decay. In this
regard, an important set of developments was the realization of the
endpoint of condensation of tachyons $localized$ to nonsupersymmetric
noncompact $\BC/\BZ_N$ and $\BC^2/\BZ_N$ orbifold singularities
\cite{aps, vafa0111, hkmm}. Since the tachyon is localized to the
singularity, its condensation can be expected to initially affect only
the immediate vicinity of the singularity: and in fact, rather than
leading to any uncontrolled behaviour, the process of tachyon
condensation smooths out and resolves the singularity. Furthermore the
physical question of tachyon condensation in these orbifold theories
is closely intertwined with the algebro-geometric structure of the
singularities \cite{hkmm} so that the known mathematics of K\"ahler
deformations can be used to study the physics of condensation of
tachyons when the condensation preserves worldsheet supersymmetry. The
corresponding question for nonsupersymmetric $\BC^3/\BZ_N$
singularities \cite{drmknmrp} is potentially somewhat more complicated
due to the existence of geometric terminal singularities (orbifolds
devoid of any marginal or relevant K\"ahler blowup modes) and the
absence of a canonical ``minimal'' resolution. Indeed a natural
question that arises is whether terminal singularities are connected
by tachyon condensation to the space of string theories. The analysis
of possible endpoints in this case can be approximated by studying the
condensation of the most relevant tachyon and sequentially iterating
this procedure for each of the residual geometries, which are
themselves typically unstable to tachyon condensation. It then turns
out that the possible $final$ endpoints for Type II string theories
(with spacetime fermions and no bulk tachyon) cannot include terminal
singularities\footnote{Note, however, that {\em canonical}\/
singularities (i.e., supersymmetric orbifold points) may be part of
the final results of tachyon condensation.} --- the combinatorics
combined with the constraints imposed by the GSO projection determines
that there is $always$ a tachyon or marginal operator in every twisted
sector of a Type II orbifold. It can however be further shown that for
Type 0 theories (with a bulk tachyon and no spacetime fermions), there
is in fact a unique truly terminal singularity $\BC^3/\BZ_2\ (1,1,1)$
(resolutions are necessarily by irrelevant operators in worldsheet
string theory at weak coupling) and it is included in the spectrum of
endpoints of tachyon condensation.

The toric geometry description used in \cite{hkmm, drmknmrp} is in
fact a reflection of the physics of an underlying gauged linear sigma
model (GLSM) of the sort first introduced by Witten
\cite{wittenphases} as a means of studying the phases of
supersymmetric Calabi-Yau spaces (and generalized to toric varieties
in \cite{morrisonplesserInstantons}). Explicit GLSM descriptions (and
mirror descriptions) of tachyons in nonsupersymmetric orbifolds have
appeared in, \eg, \cite{vafa0111, minwalla0111, sarkar0206,
martinecmoore, minwalla0307, ssin0308, ssin0312, moore0403,
sarkar0407}.

Generically, there are multiple distinct tachyons in an unstable
orbifold so that there are multiple distinct unstable channels for the
decay.  (More precisely, there is a continuous family of initial
conditions leading to a disparate set of final decay products.)  This
is particularly relevant for $\BC^3/\BZ_N$ due to the absence of
canonical ``minimal'' resolutions\footnote{By comparison, the physics
of chiral tachyon condensation dovetails beautifully with the
Hirzebruch-Jung minimal resolution formulation for $\BC^2/\BZ_N$
\cite{hkmm} ensuring a unique final K\"ahler resolution while
$\BC/\BZ_N$ is potentially too simple structurally to allow for
distinct resolutions.}: the various possible distinct resolutions are
related by flip and flop transitions which are mediated by tachyonic
and marginal operators, respectively, in the corresponding GLSM. In
this paper, we study systems of multiple tachyons and their
condensation in $\BC^3/\BZ_N$ orbifolds using GLSM descriptions of
these systems. In particular the competition between tachyons of
distinct R-charges (\ie, masses in spacetime) gives rise to flip
transitions, described in part in \cite{drmknmrp}, that relate the
distinct partial resolutions of the original singularity, thought of
as distinct basins of attraction for the worldsheet RG flow. A flip
transition can be thought of as a blowdown of a cycle accompanied by a
blowup of a topologically distinct cycle, both mediated by tachyonic
(relevant) operators. Flips can always be consistently embedded in a
2-tachyon sub-GLSM (with gauge group $U(1)^2$) of the full GLSM (with
gauge group say $U(1)^n,\ n\geq 2$), reflecting the fact that in the
corresponding toric fan, they occur in subcones representing regions
that involve ``flipping'' (reversing the sequence of subdivisions
pertaining to) only one wall in the fan.\footnote{This toric operation
is similar to the familiar ``flop'' associated to a conifold in the
supersymmetric case, but it plays a different role here both
physically and mathematically.}  Physically a flip transition occurs
for instance when a more relevant tachyon condenses during the process
of condensation of a less relevant tachyon: in this case, there turn
out to be interesting connections between the singularity structure of
the residual endpoint geometries, the combinatorics of the toric
description and the phases of the corresponding GLSM, as we will show
in what follows. We show in particular that the dynamics of the GLSM
RG flow always drives a flip transition in the direction of the
partial resolution that leads to a less singular residual geometry,
which can be thought of as a more stable endpoint. It then turns out
that the phases of GLSMs corresponding to these nonsupersymmetric
orbifolds can be classified into ``stable'' and ``unstable'' ones
noting the directionality of the RG trajectories involving potential
flip transitions, which always flow towards the more stable
phases. These ``stable'' and ``unstable'' phases of GLSMs can then be
identified with the corresponding extrema of the closed string tachyon
potential.

The ``stable'' phases survive to the infrared, and provide the
different basins of attraction to which the flow may tend.  In many
examples, there is only a single stable phase; when there are multiple
stable phases, however, they are related by marginal operators in the 
infrared.

Organization: We first review the toric description in Sec.~1. We then
describe the connections between the combinatorics, GLSMs and the
singularity structure of the residual geometries in general in Sec.~2
following it up in Sec.~3 by a detailed analysis of a Type 0 example 
containing three tachyons. In Sec.~4, we observe that
two-dimensional nonsupersymmetric orbifolds also exhibit ``stable''
and ``unstable'' phases and may also give rise to marginal operators
in the infrared. We conclude in Sec.~4 with some speculations on the 
closed string tachyon potential.

\section{Review of the toric description of $\BC^3/\BZ_N$ tachyon 
condensation}

In this section, we give a lightning review of the condensation of
closed string tachyons localized to nonsupersymmetric noncompact
$\BC^3/\BZ_N$ orbifold singularities, via renormalization group flows
that preserve supersymmetry in the worldsheet conformal field theory
and their interrelations with the combinatorial geometry of the
orbifold singularities (see \cite{drmknmrp} for details). Restricting
attention to ${\cal N}=(2,2)$ supersymmetry on the worldsheet helps us
track special classes of deformations of the geometry by studying
properties of the corresponding physical quantities which are
protected by supersymmetry. In particular, there exist various sets of
BPS-protected operators in the free conformal field theory at the
orbifold point, whose operator product expansions are nonsingular:
these constitute eight (anti-)chiral rings (in four conjugate pairs)
in the theory for codimension three, distinguished by the choices of
target space complex structure in defining the three supercurrents and
$U(1)_R$ currents and therefore the superalgebra. By convention, it is
the $(c_X,c_Y,c_Z)$ ring, chiral w.r.t.\ the orbifold coordinates
$X=z^4+iz^5,\ Y=z^6+iz^7,\ Z=z^8+iz^9$, that is called the chiral ring
of twisted states
\be\label{XcXcYcZ}
X_j \ = \ \prod_{i=1}^3\ X^{(i)}_{ \{jk_i/N\} } \ = \ X^{(1)}_{j/N}\ 
X^{(2)}_{\{jp/N\} } X^{(3)}_{\{jq/N\} },
\qquad \qquad j=1,2,\ldots N-1
\ee
constructed out of the twist fields for each of the three complex 
planes parametrized by $X,Y,Z$.

\begin{figure}
\bc
\epsfig{file=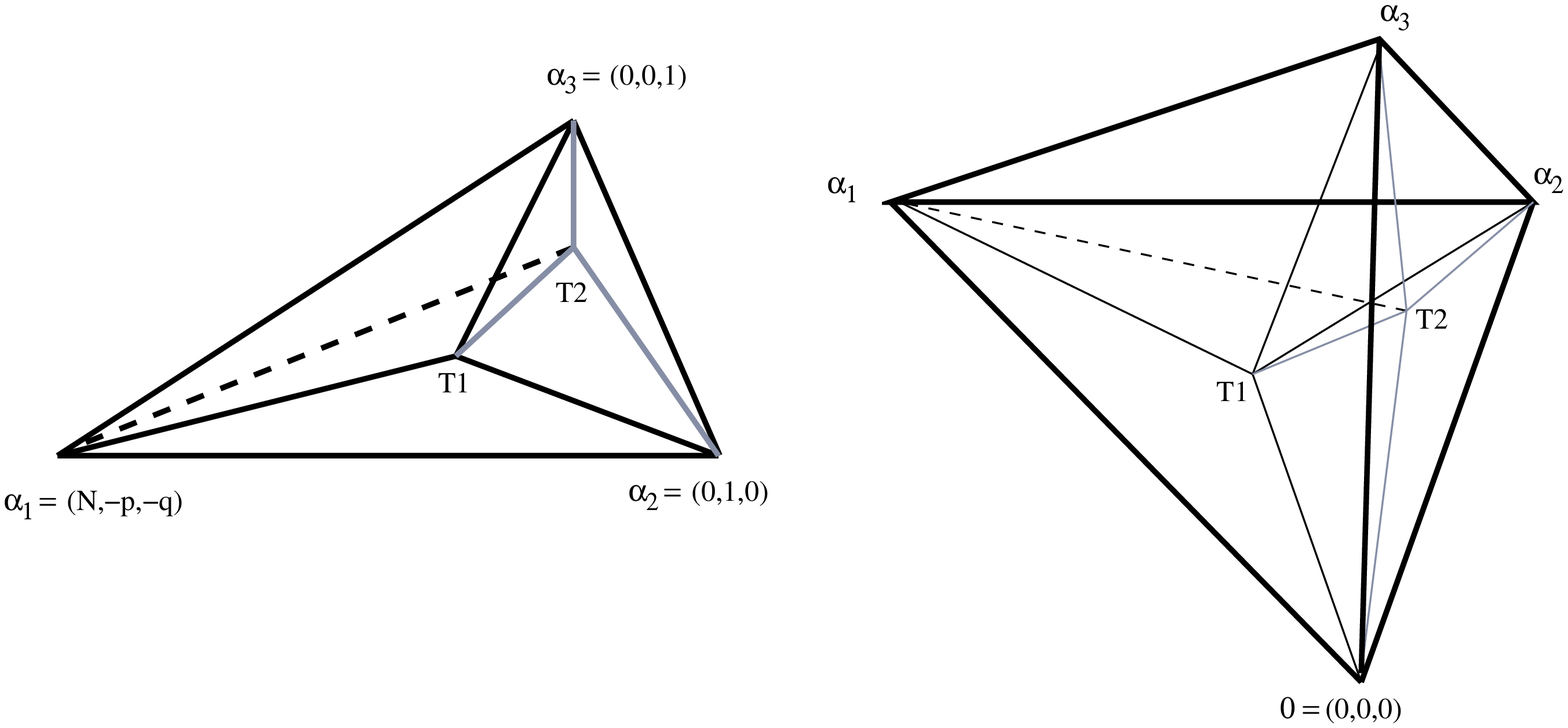, width=10cm}
\caption{The fan of cones for the $\BN$ lattice of 
$\BC^3/\BZ_{N(p,q)}$, with the vertices of the simplex $\Delta$, as 
well as tachyons $T_1, T_2$ in the interior of the cone and the 
corresponding subdivisions.
(The figure on the left shows the simplex and its subdivision; the
figure on the right shows the actual cones in the fan.)}
\label{figcone}
\ec
\end{figure}
There is a remarkable correspondence between the chiral twisted sector
states in the orbifold conformal field theory and the combinatorial
geometry of the orbifold itself. A given orbifold singularity can be
described as a (strongly convex) rational polyhedral cone in a
so-called $\BN$ lattice (see figure~\ref{figcone}): this is the
lattice of divisors (codimension one algebraic subspaces) blown down
at the singularity, dual to the $\BM$ lattice of monomials on the 
orbifold space invariant under the discrete group action. Each chiral 
twisted state can be represented uniquely as a lattice point in the
interior of this cone: \eg, the twist field $X_j$ above is the lattice
point $(j,-[{jp\over N}],-[{jq\over N}])$ in the cone. Perturbing by a
{\it chiral} twisted state in the orbifold conformal field theory
corresponds to turning on the corresponding {\it K\"ahler} blowup mode
in the geometry, \ie, blowing up the corresponding divisor.

Nonsupersymmetric orbifolds in general contain twisted sector states 
that correspond to relevant operators on the worldsheet: these describe 
tachyonic instabilities in spacetime (\ie, the unorbifolded dimensions) 
with masses 
\be
M^2 = {4\over \al'} \bigg( h - {1\over 2} \bigg) < 0\ ,
\ee 
where $h={|q|\over 2}$ is the conformal weight with $|q|<1$ being the
R-charge of the relevant operator. Then a perturbation by such a
relevant operator induces a renormalization group flow from the
ultraviolet fixed point, \ie, the unstable orbifold, to infrared fixed
points: the divisor corresponding to the twisted state blows up giving
rise to a partial resolution of the singularity. The expanding divisor
locus typically contains residual singularities that are decoupled in
spacetime from each other in the large radius limit, \ie, the infinite
RG-time limit. The combinatorics of the geometry coupled to the fact
that we are restricting attention to worldsheet-supersymmetry
preserving deformations enables us to follow the possible 
renormalization group flows all the way to their endpoints. Furthermore 
for Type II string theories, an appropriate GSO projection can be
consistently imposed: the combinatorics along with this GSO projection
is sufficient to show the nonexistence in codimension three of
potential {\it terminal} singularities (orbifolds devoid of any blowup
modes, K\"ahler or otherwise) thus implying that the endpoints of
localized closed string tachyon condensation for Type II theories are
always supersymmetric spaces\footnote{That is, any remaining
singularities are of the type which preserves supersymmetry.}  in
codimension three as well \cite{drmknmrp}, just as in the lower
codimension cases \cite{aps, vafa0111, hkmm}. For Type 0 theories
however, the combinatorics does show the existence of a unique
all-ring terminal singularity, $\BC^3/\BZ_2\ (1,1,1)$, which is
included in the endpoints of tachyon condensation \cite{drmknmrp}.

One of the new features in codimension three described in
\cite{drmknmrp} was the appearance of flip transitions when multiple
tachyons of differing degrees of tachyonity, \ie, different worldsheet
R-charges (or masses in spacetime), condense simultaneously. In such
cases, there are potentially multiple basins of attraction so that in
intermediate stages, distinct condensation endpoints are possible: in
particular, it was found there that a more relevant tachyon gives rise
to a less singular partial resolution endpoint and vice versa.
Furthermore the renormalization group trajectories corresponding to a
less relevant tachyon can be forced to turn around if the more
relevant tachyon turns on during this process: this turn-around
behaviour is a flip transition. In this work, we describe the physics
of these transitions, using explicit gauged linear sigma model
constructions of the toric descriptions in \cite{drmknmrp}.
These constructions allow us to study the RG flow for arbitrary initial
expectation values of the condensing tachyons.

\begin{figure}
\bc
\epsfig{file=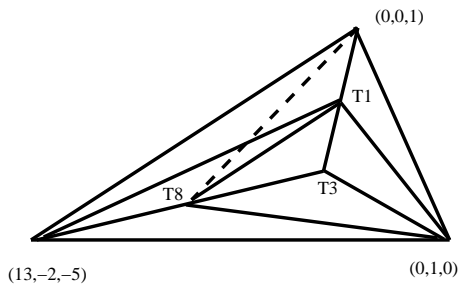, width=6cm}
\caption{$\BC^3/\BZ_{13}\ (1,2,5)$ : The three points defining the 
affine hyperplane $\Delta$ of marginal operators are shown, along with 
the three tachyons $T_1, T_3, T_8$ and two distinct sequences of 
subdivisions. The solid lines correspond to the sequence of most 
relevant tachyons.}
\label{fig1}
\ec
\end{figure}
To facilitate the sections to follow, we briefly review here the toric 
description of an example discussed in \cite{drmknmrp} (see 
figure~\ref{fig1}) which we will study in later sections using GLSMs.\\

{\footnotesize
{\bf Example $\BC^3/\BZ_{13}\ (1,2,5)$} : In the Type 0 theory, there 
are three tachyons $T_1, T_3, T_8$ in the $(c_X,c_Y,c_Z)$ ring with 
R-charges $R_1=({1\over 13},{2\over 13},{5\over 13})={8\over 13},\ 
R_3=({3\over 13},{6\over 13},{2\over 13})={11\over 13}$ and 
$R_8=({8\over 13},{3\over 13},{1\over 13})={12\over 13}$, respectively. 
The most relevant tachyon in the Type 0 theory is in fact $T_1$ above. 
In the toric fan (with vertices\ $\al_1=(13,-2,-5),\ \al_2=(0,1,0),\ 
\al_3=(0,0,1)$), these tachyons correspond to the lattice vectors 
$T_1=(1,0,0),\ T_3=(3,0,-1),\ T_8=(8,-1,-3)$. It is possible to show 
using the combinatorics of the geometry that $T_1$ and $T_3$ are 
coplanar with $\al_3$ while $T_3$ and $T_8$ are coplanar with $\al_1$. 
Let us first consider the subdivisions shown by the solid lines which 
correspond to sequentially blowing up the most relevant tachyon, \ie, 
the sequence $T_1,\ T_3,\ T_8$: this gives the total $\BN$ lattice 
volume of the subcones, \ie, the degree of the residual singularity, to 
be $6(1)+2$. Let us consider this sequence in more detail: the first 
subdivision by $T_1$ gives the three decoupled residual geometries as 
the three subcones $C(0;\al_3,T_1,\al_2),\ C(0;\al_1,T_1,\al_3),\ 
C(0;\al_1,\al_2,T_1)$, which can be shown to respectively be flat 
space, the terminal singularity $\BZ_2\ (1,1,-1)$, and the 
supersymmetric $\BZ_5\ (1,2,-3)$ orbifold with marginal blowup modes 
(no relevant blowup modes). The renormalized R-charges of the 
subsequent twisted states can be shown to be $R_3'=1,\ R_8'=1$.

On the other hand, choosing a different sequence of tachyons by which
to fully subdivide the cone gives rise to different fans of subcones.
For instance, the dotted line in figure~\ref{fig1} shows the
subdivision corresponding to the sequence $T_3,\ T_8,\ T_1$ and the
corresponding flip transition. From the combinatorics, it can be shown
that the subdivision by $T_3$ gives the subcones
$C(0;\al_2,\al_3,T_3),\ C(0;T_3,\al_3,\al_1)$ and 
$C(0;\al_1,\al_2,T_3)$, to be $\BZ_3\ (0,1,1),\ \BZ_6\ (13,-2,-3)\equiv 
\BZ_6\ (1,-2,-3)$ and $\BZ_2\ (-3,-6,13)\equiv \BZ_2\ (1,0,1)$ 
singularities, respectively. Further it is straightforward to show that 
$T_8$ becomes marginal after the blowup by $T_3$ while $T_1$ acquires 
the R-charge $R_1'={2\over 3}>{8\over 13}$. One can now subdivide by 
the remaining relevant tachyons to obtain the total volume of the 
subcones to be $6(1)+3$ : this is greater than the corresponding total 
volume in the most-relevant-tachyon subdivision. Thus if the tachyons 
$T_1, T_8$ condense simultaneously, then the tachyon dynamics would 
appear to drive the system towards the less singular endpoint, which is 
given by the most-relevant-tachyon sequence. In what follows, we will 
flesh this out in greater detail. }

\section{Phases of GLSMs}

\subsection{The condensation endpoint of a single tachyon}

The condensation of a single tachyon can be modeled \cite{vafa0111} by
a gauged linear sigma model (GLSM) in two dimensions with $(2,2)$
worldsheet supersymmetry with a gauge group $U(1)$ and chiral
superfields $\Psi_i\equiv \phi_1, \phi_2, \phi_3, T$.  Following the
conventions of \cite{wittenphases, morrisonplesserInstantons}, we
describe the charges under the $U(1)$ action by a charge vector
\be\label{Qia1}
Q_i = (r_1, r_2, r_3, -N) ,
\ee
\ie, these fields transform under $U(1)$ gauge transformations as 
\be\label{U1gt}
\Phi_i \ra e^{i r_i\lambda} \Phi_i, \qquad T \ra e^{-iN\lambda} T .
\ee
The action for the GLSM (with no chiral superpotential, \ie, F-terms) 
is \be
S = \int d^2 z\ \biggl[ d^4 \theta\ \biggl( {\bar \Psi_i} e^{2Q_i V} 
\Psi_i - {1\over 4e^2} {\bar \Sigma} \Sigma \biggr) + \Rea\biggl( i 
t\int d^2 {\tilde \theta}\ \Sigma  \biggr) \biggr]\ ,
\ee
where $t = ir + {\theta\over 2\pi}$, with $\theta$ the $\theta$-angle 
in $1+1$-dimensions and $r$ a Fayet-Iliopoulos (FI) parameter that 
plays a nontrivial role in determining the vacuum structure of the 
theory.\footnote{Note that since the $\theta$-angle is only determined
up to multiples of $2\pi$, the physically relevant parameter is actually
$e^{2\pi it}$.  In particular, it makes sense to take $r\to\infty$, which 
simply sends $e^{2\pi it}$ to $0$.}
The classical vacua of the 2D theory that preserve supersymmetry are
given by solving the F- and D-term equations for the system: since
there is no superpotential, there are no F-term equations, and the
single D-term equation gives
\be
{-D\over e^2} = \sum_j q_j |\psi_j|^2 - r = \sum_{i=1}^3 r_i |\phi_i|^2 
- N |T|^2 - r = 0 .
\ee

Now as we vary the parameter $r$, the vacuum structure of the system
changes. For $r \ll 0$, it is straightforward to see that this D-term
equation has a solution only if the field $T$ has nonzero vacuum
expectation value. Thus, the modulus of the expectation value of the
field $T$ is fixed, and this Higgses the $U(1)$ down at low energies
(relative to the scale set by $e$) to a residual discrete $\BZ_N$
gauge group with generator $\omega=e^{2\pi i/N}$, as can be seen from
the gauge transformations (\ref{U1gt}). This residual $\BZ_N$ acts on
the $\Phi_i$ fields as $\ \Phi_k \ra e^{2\pi i r_k/N} \Phi_k$. The
action for fluctuations of the fields $\Phi_i$ with these restricted
gauge transformations in fact yields the action for a nonlinear sigma
model on a $\BC^3/\BZ_N\ (r_1,r_2,r_3)$
orbifold\footnote{Ref.~\cite{sarkar0407} computes the sigma model
metric for $\BC^2/\BZ_N$ and $\BC^3/\BZ_N$.}  with the coordinate
chart $(\phi_1, \phi_2, \phi_3)$.  On the other hand, for $r \gg 0$,
we can similarly see that the D-term equation admits a solution only
if at least one of the fields $\Phi_i$ has a nonzero vacuum
expectation value with scale $O(r)$. For instance, the D-term equation 
simplifies for $|\phi_i| \gg |T|$ to
\be
r_1 |\phi_1|^2 + r_2 |\phi_2|^2 + r_3 |\phi_3|^2 = r \gg 0
\ee
which describes a weighted $\BC \BP^2$ with K\"ahler class $r$ in 
$\Phi$-space. More generally, the D-term equation describes the divisor 
corresponding to the tachyonic operator in the toric description. In 
greater detail, in the region of moduli space where, say, $\phi_3 \neq 
0$ with the other expectation values vanishing, the $U(1)$ gauge group 
is broken down to $\BZ_{r_3}$, with action $\Phi_i \ra e^{2 \pi i 
r_i/r_3} \Phi_i , \ i=1,2, \ T \ra e^{-2\pi i N/r_3} T$: the remaining 
massless fields yield a description of a space with the coordinate 
chart $(\phi_1, \phi_2, T)$ characterizing the $\BC^3/\BZ_{r_3}$ 
orbifold singularity. Similarly the regions of moduli space where 
$\phi_1$ or $\phi_2$ alone acquires an expectation value yield 
descriptions of spaces with coordinate charts $(\phi_2, \phi_3, T)$ and 
$(\phi_1, \phi_3, T)$ characterizing $\BC^3/\BZ_{r_1}$ and 
$\BC^3/\BZ_{r_2}$ singularities, respectively. The regions of moduli 
space where multiple fields acquire expectation values describe 
overlaps of these coordinate charts. The collection of coordinate 
charts $\{\ (\phi_1, \phi_2, T),\ (\phi_2, \phi_3, T),\ (\phi_1, 
\phi_3, T)$\ \} characterizes a toric variety describing the blown-up 
divisor with residual $\BC^3/\BZ_{r_i}$ singularities. The presence of 
$(2,2)$ worldsheet supersymmetry ensures that the superHiggs mechanism 
preserves precisely the light fields required for the above geometric 
descriptions for each of these $r$ values.

We see that the region $r \ll 0$ describes the unresolved
$\BC^3/\BZ_N\ (r_1,r_2,r_3)$ orbifold described by coordinates
represented by the fields $\Phi_i$. The field $T$ is the chiral
primary representing the tachyon with R-charges 
$({r_1\over N},{r_2\over N},{r_3\over N})$: we have 
$R_T = \sum_i {r_i\over N} < 1$ 
since $T$ is tachyonic. On the other hand, the region $r \gg 0$
describes the endpoint of condensation of the tachyon $T$ executing
the partial resolution of the $\BC^3/\BZ_N\ (r_1,r_2,r_3)$
singularity, with residual $\BC^3/\BZ_{r_i}$ singularities. Thus the
target spacetime geometry emerges as the moduli space of the GLSM in
its classical phases, changing nontrivially depending on the FI
parameter $r$ in the GLSM.

The quantum story is rich. The parameter $r$ receives 1-loop 
corrections \be
r \equiv r(\mu) = \sum_i Q_i\cdot\ \log {\mu\over \Lambda} = 
\biggl(\sum_i r_i - N\biggr)\cdot\ \log {\mu\over \Lambda} = N (R_T - 
1)\cdot\ \log {\mu\over \Lambda}\ .
\ee
This is related to a potential anomaly in the $U(1)_R$ symmetry for 
$\sum_i Q_i\neq 0$. In this 1-loop correction, $r$ is renormalized to 
vanish at the energy scale $\Lambda$. By the usual nonrenormalization 
theorems of $(2,2)$ supersymmetry in two dimensions, there are no 
further perturbative corrections. While nonperturbative (instanton) 
corrections are in general present, they will be negligible for our 
discussions here since we will largely be concerned only with the 
physics of the large $|r|$ regions. There are different qualitative 
behaviors of the nature of quantum corrections to $r$, depending on 
the value of $\sum_i Q_i$.  If $\sum_i Q_i = 0$, then there are no
1-loop corrections and $r$ is a marginal parameter for large $|r|$. On
the other hand, for a tachyonic field $T$ with the charge vector
(\ref{Qia1}), we have $\sum_i Q_i = N(R_T - 1) < 0$. Thus the 1-loop
correction is nonzero and induces a renormalization group flow: $r$
flows from the region $r \ll 0$ (orbifold phase) at high energies $\mu
\gg \Lambda$, to $r \gg 0$ (partial resolution by $T$) at low energies
$\mu \ll \Lambda$. For $\sum_i Q_i > 0$, the RG flow is reversed in
direction, corresponding to an irrelevant operator.

We mention here that since the gauge coupling $e$ in two dimensions has 
mass dimension one, energy scales in any physical process are defined
relative to the scale set by $e$. Thus there are two scales in the
system, $e$ and $\Lambda$ (the scale set by $r$), and they are
independent. The full RG flow in the GLSM has two components: the
first from free gauge theory (free photons) in the ultraviolet
(energies much larger than the gauge coupling $e$) to the infrared
(energies small compared to $e$), the second entirely in the IR, at
low energies relative to $e$, but with nontrivial dynamics relative to
the scale $\Lambda$, described in the preceding paragraph. At energies
much lower than $e$, fluctuations transverse to the space of vacua 
(moduli space) cost a lot of energy. Thus the low-lying fluctuations 
are simply scalars defining the moduli space of the theory so that the
possibility of a geometric description emerges, given by a nonlinear
sigma model on the moduli space.

It is noteworthy that the RG flows in the GLSM are in principle
distinct from those in the nonlinear sigma model. The $endpoints$ of
the RG flows in the GLSM being classical phases coincide with those of
the nonlinear model and the GLSM RG flows themselves approximate the
nonlinear ones in the low-energy regime $e^2 \ra \infty$. The
$(2,2)$-supersymmetry preserving RG flows of the GLSM can be reliably
tracked in the corresponding topologically twisted A-model: the
topological A-twist retains information about K\"ahler deformations of 
the model while complex structure information is in general lost. Since 
tachyon condensation corresponds to K\"ahler deformations of the 
orbifold, topological twisting is a sufficient tool to understand the 
physics of the situation. Since attention is restricted to 
quasi-topological observables, the gauge-coupling $e^2$ itself is not 
crucial to the discussion.

\subsection{Flip transitions and $\BN$ lattice volume 
minimization}

Consider a GLSM with gauge group $U(1)\times U(1)$ and two of five 
chiral superfields representing tachyons $T_1$ and $T_2$ with R-charges 
$R_1\equiv ({a_1\over N},{b_1\over N},{c_1\over N})
={a_1\over N}+{b_1\over N}+{c_1\over N}$ and $R_2\equiv ({a_2\over 
N},{b_2\over N},{c_2\over N})
={a_2\over N}+{b_2\over N}+{c_2\over N}$. The action of the $U(1)\times 
U(1)$ on the scalar fields $\phi_1, \phi_2, \phi_3, T_1 T_2$ is given 
by the charge matrix \be\label{Qia2gen}
Q_i^a = \left( \bA{ccccc} a_1 & b_1 & c_1 & -N & 0  \\ a_2 & b_2 & c_2 
& 0 & -N  \\ \eA \right), \qquad i=1,\ldots,5, \ \ a=1,2\ .
\ee
(Such a charge matrix only specifies the $U(1)\times U(1)$ action up to 
a finite group, due to the possibility of a $\BQ$-linear combination of 
the rows of the matrix also having integral charges.)

There are two independent FI parameters (for the two $U(1)$'s)
whose variations control the vacuum structure of the theory. The
classical vacua of the theory can be again found by studying the
D-term equations
\be
{-D_i\over e^2} = a_i |\phi_1|^2 + b_i |\phi_2|^2 + c_i |\phi_3|^2 - N 
|T_i|^2 - r_i = 0\ , \qquad i=1,2\ .
\ee
This 2-parameter system admits several ``phases'' depending on the
values of $r_1, r_2$ and we shall analyze such a system in detail in
what follows.  For the moment, let us focus attention on the region of
moduli space where the nontrivial dynamics does not involve the field
$\phi_2$: in other words, we look for a linear combination of the two
$U(1)$'s which does not include $\phi_2$ and study the effective
dynamics of this subsystem. In this case, the linear combination 
$b_2 A_1 - b_1 A_2$ of the two gauge fields can be seen to not couple
to $\phi_2$: the corresponding D-term is
\bea
{-D^{eff}\over Ne^2} &=& {1\over N} (b_2 D_1 - b_1 D_2) \nonumber\\
&=& {1\over N} (a_1 b_2 - a_2 b_1) |\phi_1|^2 + {1\over N} (c_1 b_2 - 
c_2 b_1) |\phi_3|^2 - b_2 |T_1|^2 + b_1 |T_2|^2 - r^{eff} ,
\eea
so that the dynamics of this effective 4-field GLSM is governed by\ 
$r^{eff} = {1\over N} (b_2 r_1 - b_1 r_2)$ . The detailed physics of 
this system strongly depends on the combinatorics encoded by the charge 
matrix $Q_i^a$. For instance, in the example $\BC^3/\BZ_{13}\ (1,2,5)$ 
studied in Sec.~3, the corresponding $D^{eff}$-term equation, using the 
charge matrix (\ref{Qia2param}), is 
\be
{-D^{eff}\over Ne^2} = |\phi_3|^2 + 2|T_2|^2 - |\phi_1|^2 - 3|T_1|^2 - 
r^{eff} = 0\ .
\ee
Notice that this effective D-term has the same schematic form as that 
for a supersymmetric conifold exhibiting a flop\footnote{Recall that a 
supersymmetric conifold can be expressed as a hypersurface in $\BC^4$: 
then in terms of a $U(1)$ GLSM with fields $z_1, z_2, z_3, z_4$ 
carrying charges $(+1,+1,-1,-1)$, respectively, under the $U(1)$, the 
corresponding D-term is 
\bea
{-D^{flop}\over e^2} &=& |z_1|^2 + |z_2|^2 - |z_3|^2 - |z_4|^2 - \rho\ . 
\nonumber
\eea
The conifold singularity itself is at $\rho=0$ classically. Redefining 
the variables $z_i$ can be used to recast the above equation in perhaps 
more familiar forms.}, except that under the $U(1)$ in question here, 
the fields $\phi_1, \phi_3, T_1, T_2$ carry charges 
\be
Q_i = \left( \bA{cccc} {1\over N} (a_1 b_2 - a_2 b_1) & {1\over N} (c_1 
b_2 - c_2 b_1) & -b_2 & b_1 \eA \right)\ \ra\ 
\left( \bA{cccc} 1 & 2 & -1 & -3 \eA \right)\ ,
\ee
so that $\sum_i Q_i \neq 0$. This is a reflection of tachyons at play,
giving an inherent directionality to the dynamics. This is a
nonsupersymmetric analogue of a conifold region, with nontrivial
dynamics in time. We can study the phases of this effective 4-field
GLSM along the same lines as the previous section, and along the same
lines as for the flop described \eg, in \cite{wittenphases} 
\cite{wittenIAS}.\\
For $r^{eff}\ll 0$, we have $|\phi_1|^2 + 3|T_1|^2 \sim -r^{eff}$, 
describing a $\mathbb{P}^1_-$ with residual singularities described by 
the coordinate charts $(\phi_3,T_2,\phi_1), \ (\phi_3,T_2,T_1)$. For 
$r^{eff}\gg 0$, we have $|\phi_3|^2 + 2|T_2|^2 \sim r^{eff}$, 
describing a $\mathbb{P}^1_+$ with residual singularities described by 
the coordinate charts $(\phi_3,\phi_1,T_1), \ (T_2,\phi_1,T_1)$.

Now from the 1-loop running of the FI parameters, we have 
\be\label{r1r21loop}
r_1(\mu) = N (R_1 - 1)\cdot \log {\mu \over \Lambda} , \qquad \qquad
r_2(\mu) = N (R_2 - 1)\cdot \log {\mu \over \Lambda} , \ee
so that the 1-loop running of $r^{eff}$ is 
\be\label{reff}
r^{eff} = \biggl[ \biggl({a_1\over N} + {c_1\over N} - 1\biggr) b_2
- \biggl({a_2\over N} + {c_2\over N} - 1\biggr) b_1 \biggr]\cdot \log 
{\mu \over \Lambda}\ \ra\ (-1)\ \log {\mu \over \Lambda}\ .
\ee

Thus $r^{eff}$ flows under the renormalization group from the 
$r^{eff}\ll 0$ phase (partial resolution by $\mathbb{P}^1_-$) to the
$r^{eff}\gg 0$ phase (partial resolution by $\mathbb{P}^1_+$). This is 
the GLSM reflection of the flip transition in spacetime -- 
clearly it involves the blowdown of the 2-cycle $\mathbb{P}^1_-$ 
followed by a blowup of the topologically distinct $\mathbb{P}^1_+$. 
\begin{figure}
\bc
\epsfig{file=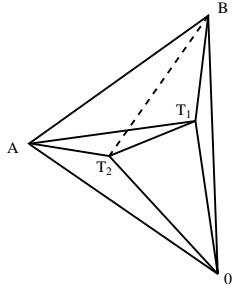, width=3cm}
\caption{A flip region in the toric fan for 
$\BC^3/\BZ_{13}\ (1,2,5)$ defined by the convex hull $\{ A, B, T_1, T_2 
\}$. The solid lines show the subdivision for the sequence $T_1, T_2$ 
giving smaller $\BN$ lattice volume while the dotted line shows the 
subdivision for the reversed sequence.}
\label{figflip}
\ec
\end{figure}
To understand the significance of the coefficient of the logarithm in 
(\ref{reff}), it is helpful to revisit the toric description of the 
orbifold decay \cite{drmknmrp} and the corresponding flip region.
Condensation of a localized tachyon gives rise to an expanding bubble
(containing a divisor) 
with three decoupled regions which can potentially 
contain residual singularities. We now review the relation 
\cite{drmknmrp} between the relevance (R-charges) of distinct tachyons 
and the $\BN$ lattice volumes of the subcones (which effectively gives 
the cumulative order or degree of the residual singularities) resulting 
from the subdivisions corresponding to those tachyons. Let us begin 
with the cone $C(0;\al_1,\al_2,\al_3)$ corresponding to the unresolved 
$\BC^3/\BZ_{N(p,q)}$ singularity (see figure~\ref{figcone}). Now add a 
lattice point $T_j=(j,-[{j p\over N}],-[{j q\over N}])$ corresponding 
to a tachyon $T_j$. Then the total $\BN$ lattice volume of the three 
residual subcones after subdivision with the tachyon $T_j$ is easily 
calculated: let $V(e_0;e_1,e_2,e_3)$ denote the volume of a cone 
subtended by the vectors $e_1,e_2,e_3$ with the vector $e_0$ being the 
apex. Then (see, \eg, figure~\ref{figcone}) we have 
\be
V_{subcones}(T_j) = V(0;\al_1,\al_2,\al_3) - V(T_j;\al_1,\al_2,\al_3)
= N R_j < N\ ,
\ee
where the volume of a cone $V(0;e_1,e_2,e_3) = |{\rm det} 
(e_1,e_2,e_3)| = |e_1\cdot (e_2\times e_3)|$. Now if we consider two 
tachyons $T_1$ and $T_2$ (as in figure~\ref{figcone}), such that $T_1$ 
is more relevant than $T_2$, \ie,  $R_1 < R_2$, where $R_i = {j_i\over 
N} + \{ {j_i p \over N} \} + \{ {j_i q \over N} \}$, this implies\ 
$V_{subcones}(T_1) < V_{subcones}(T_2)$.
Thus given an orbifold singularity, isolated condensation of a more 
relevant tachyon locally leads to a less singular partial resolution.
Now if both tachyons simultaneously condense, then the toric
description shows the existence of a region exhibiting a flip
transition. For instance, in $\BC^3/\BZ_{13}\ (1,2,5)$, the region in
question is given by the convex hull subtended by the $\BN$ lattice
points $\al_1, T_8, T_1, \al_3$ with the origin $(0,0,0)$ (see
figure~(\ref{fig1})). The most-relevant-tachyon sequence $T_1, T_8,
T_3$ (\ie, the subdivision given by the solid lines) results in a less
singular endpoint geometry than, say, the sequence $T_8, T_1, T_3$. In
the general case we describe above, the corresponding flip region is
the convex hull subtended by the $\BN$ lattice points $A \equiv
(N,-p,-q), \ \ B \equiv (0,0,1), \ \ T_1 \equiv (j_1,-[{j_1 p\over
N}],-[{j_1 q\over N}]), \ \ T_2 \equiv (j_2,-[{j_2 p\over N}],-[{j_2
q\over N}])$\ \ with the origin $0\equiv (0,0,0)$, as shown in
figure~\ref{figflip}. We can calculate the difference in the $\BN$
lattice volumes of the corresponding subcones to get a quantitative
measure of the difference in the degree of singularity between the two
blowup sequences as
\be
\Delta V = V(0;A,T_1,B) + V(0;A,T_2,T_1) - V(0;B,T_2,T_1) + 
V(0;A,T_2,B)\ .
\ee
Using the coordinate values for the lattice points, the expression 
above can be simplified to give 
\be
\Delta V = \biggl( {j_1\over N} + \biggl\{ {j_1 q\over N} \biggr\} - 1 
\biggr)\ N \biggl\{ {j_2 p\over N} \biggr\} - \biggl( {j_2\over N} + 
\biggl\{ {j_2 q\over N} \biggr\} - 1 \biggr)\ N \biggl\{ {j_1 p\over N} 
\biggr\}\ .
\ee
By rewriting this expression in terms of the R-charges $(a_i,b_i,c_i)$ 
for the tachyons $T_1, T_2$, it is straightforward to recognize (see 
eqn.(\ref{reff})) that this is precisely the coefficient of the 
logarithm in $r^{eff}$, \ie, 
\be
r^{eff} = (\Delta V)\cdot \log {\mu \over \Lambda}\ .
\ee
Thus the RG flow for this effective FI parameter proceeds precisely in 
the direction of decreasing $\BN$ lattice volume, \ie, in the direction 
along which $\Delta V < 0$. In other words, the renormalization group 
dynamics in the GLSM drives the flip transition in the direction of 
the partial resolution that leads to a less singular residual geometry, 
which can be thought of as a stable endpoint of this effective RG flow.

If the tachyon $T_1$ is more relevant than $T_2$, then we have $R_1 < 
R_2$: the 1-loop coefficient in the FI parameter for a given tachyon is 
related to its R-charge. As we have seen, when the tachyons condense 
individually, the cumulative volume of the residual singularities is 
smaller for $T_1$, in other words, the more relevant tachyon leads to a 
less singular endpoint and vice versa. What the above calculation shows 
is that when two tachyons condense simultaneously, the 1-loop 
coefficient in the relative FI parameter $r^{eff}$ is related to the 
difference in the degrees of the residual singularities, \ie, the $\BN$ 
lattice volume difference between the two toric subdivisions performed 
in either of the two possible orders.

\subsection{The $n$-tachyon system: some generalities}

We will now describe some generalities\footnote{Due to the possibly 
technical nature of this subsection, it might be useful to read this in 
conjunction with the next section which elucidates these generalities 
in a specific example.} on closed string tachyon condensation in an 
unstable orbifold with multiple tachyons via gauged linear sigma 
models\footnote{We continue to use conventions of \cite{wittenphases, 
morrisonplesserInstantons}.}. The action for a GLSM for a target space 
$\BC^d/\BZ_N$ can be written as 
\be
S = \int d^2 z\ \biggl[ d^4 \theta\ \biggl( {\bar \Psi_i} e^{2Q_i^a 
V_a} \Psi_i - {1\over 4e_a^2} {\bar \Sigma_a} \Sigma_a \biggr) + 
\Rea\biggl( i t_a\int d^2 {\tilde \theta}\ \Sigma_a  \biggr) \biggr]\ ,
\ee
where summation on the index $a=1,\ldots, n$ is implied. The\ $t_a = 
ir_a + {\theta_a\over 2\pi}$ \ are Fayet-Iliopoulos parameters and 
$\theta$-angles for each of the $n$ gauge fields. The twisted chiral 
superfields $\Sigma_a$ (whose bosonic components are complex scalars 
$\sigma_a$) represent field-strengths for the gauge fields. The various 
chiral superfields $\Psi_i,\ i=1,\ldots,d+n$ in the GLSM are the $d$ 
coordinate superfields $\Phi_i$ and the $n$ tachyons $T_k$ : for $d=3$, 
the action of the $U(1)^n$ gauge group on them is given as 
\be\label{Qiagen}
\Psi_i \ra e^{i Q_i^a\lambda}\ \Psi_i\ , \qquad Q_i^a = \left( 
\bA{ccccccc} q_1^1 & q_2^1 & q_3^1 & -N & 0 & \ldots \\ q_1^2 & q_2^2 & 
q_3^2 & 0 & -N & \ldots  \\ & & \cdot & & & \ldots  \\ & & \cdot & & & 
\ldots  \\ \eA \right), \qquad \ a=1,\ldots,n\ 
\ee
where $Q_i^a$ is the $n\times (d+n)$ charge matrix
(although in order to accurately specify the group action, some of these
rows may need to replaced by $\BQ$-linear combinations of the original 
rows). From the point of view of the target spacetime, this gives the 
R-charges of the $n$ tachyons to be $R_a\equiv ({q_1^a\over 
N},{q_2^a\over N},{q_3^a\over N}) = \sum_{i=1}^3 {q_i^a\over N}\ , \ 
a=1,\ldots,n$. The theory described by this GLSM is 
super-renormalizable: the only divergence arises in the 1-loop 
renormalization of the $n$ Fayet-Iliopoulos parameters, given by 
\be
r_a = \bigg( \sum_i Q_i^a \bigg)\cdot \log {\mu\over \Lambda}\ ,
\ee
where $\mu$ is the RG scale and $\Lambda$ is an ultraviolet cutoff 
scale where the $r_a$ are defined to vanish. Thus all RG flowlines in 
$r$-space are straight lines at the level of these 1-loop equations. In 
general there may be nonperturbative (instanton) corrections to the
theory in the small $r$ regions, which are expected to transform these
straight flowlines into curves: for our purposes here, we will not
need to consider these since we will only be interested in the phases 
of the GLSM for large $r$. If we consider GLSMs with strictly relevant 
operators, we have $\sum_i Q_i^a < 0$ for all $a=1,\ldots,n$: then the 
GLSM renormalization group drives the system towards the region where 
all of the $r_a > 0$.

The space of classical ground states of this theory can be found by the 
bosonic potential 
\be
U = \sum_{a=1}^n {(D_a)^2\over 2e_a^2} + 2\sum_{a,b=1}^n {\bar 
\sigma}_a \sigma_b \sum_{i=1}^{n+d} Q_i^a Q_i^b |\phi_i|^2\ .
\ee
Then $U=0$ requires $D_a=0$: solving these for $r_a=0$ gives 
expectation values for the $\phi_i$ which Higgs the $U(1)^n$ down to 
$Z_N$ and lead to mass terms for the $\sigma_a$ whose expectation 
values thus vanish. The classical phase diagram of the GLSM is then 
given in terms of solutions to the D-term equations \be
-{D_a\over e_a^2} = \sum_{i=1}^{d+n} Q_i^a |\psi_i|^2 - r_a = 0\ , 
\qquad a=1,\ldots,n\ .
\ee
These give collections of coordinate charts that characterize in 
general distinct toric varieties, depending on the $n$ FI parameters 
$r_a$. In other words, as we vary $r_a$ we pass through several 
distinct ``phases'' \cite{wittenphases} in $r$-space that represent in 
general distinct target space geometries. The explicit description of 
these toric varieties is a standard exercise in toric geometry, which 
we now review.

We have studied the 2-tachyon $\BC^3/\BZ_N$ system in the previous
subsection on flip transitions and will study it in greater detail
later.  Specialize now to $\BC^3/\BZ_N$ and $n>2$: then we find in
$r$-space (which is $n$-dimensional here) several distinct phase
bounding $n$-vectors $\phi_1\equiv (q_1^1,q_1^2,\ldots) , \
\phi_2\equiv (q_2^1,q_2^2,\ldots) , \ \phi_3\equiv
(q_1^3,q_3^2,\ldots) , \ \ T_1\equiv (-N,0,\ldots) ,\ T_2\equiv
(0,-N,\ldots) ,\ \ldots$: these can be read off as the $3+n$ columns
in the charge matrix (\ref{Qiagen}). Hyperplanes of various
dimensionalities embedded in $r$-space subtended by these phase
bounding vectors give rise to phase boundaries. At the location of
such a phase boundary, some (but not all) of the $r_a$ vanish thus
giving rise to singularities classically. These hyperplanes defining
phase boundaries intersect along rays emanating from the origin in
$r$-space.  The set of these intersection rays includes the $3+n$
vectors $\psi_i\equiv \phi_i, T_j$ but also in general contains new
vectors $\{ \phi_{Int} \}$. Each classical phase of the GLSM is thus
an $n$-dimensional convex hull (or $n$-simplex) in $r$-space, \ie, the
interior region of $n$-tuples of the full set of intersection rays $\{
\phi_i, T_j, \phi_{Int} \}$. We can gain insight into the geometric
structure of $r$-space by gleaning the geometric relations amongst the
intersection rays, effectively treating them as toric data and
studying their combinatorics: doing so results in the {\it secondary
fan} of the system.

Note that a given convex hull may itself be contained in other convex 
hulls: thus a given region in $r$-space typically admits more than 
simply one description on the moduli space of the GLSM in terms of 
coordinate charts in spacetime. Thus each convex hull or phase (which 
is a well-defined region in $r$-space) yields a collection of 
coordinate charts in general (rather than a single chart), which then 
describes a toric variety in spacetime.

It is important to note that $k$-tachyon subsystems of this
$n$-tachyon system can be studied consistently independent of the
other $n-k$ tachyons by taking appropriate limits of the original
GLSM. The moduli space of the GLSM for the subsystem is given as the
subspace obtained by restricting to $r_\ell\ra -\infty,\
\ell=k+1,\ldots,n-k$. This decoupling of the $n-k$ tachyons is
consistent, showing that condensation of some of the tachyons does not
source the remaining ones\footnote{This is consistent with the results
of \cite{atish0403} for $\BC/\BZ_N$ who compute off-shell correlators
for various tachyons (twisted states) and find that the cubic
interaction vanishes within their approximations, showing that
condensation of one tachyon does not source others.}: we will have
more to say on this later when we describe in detail the stability of
the phases of the GLSM in a specific example.

It is convenient to choose a basis for $Q_i^a$ such that $a=1$ 
represents the most relevant tachyon. Then we have\ $R_1 \leq R_b,\ 
b\neq 1$\ so that\ $|\sum_i Q_i^1| = |N (R_1 - 1)| \geq |\sum_i 
Q_i^b|,\ b\neq 1$. A generic linear combination of the $n$ $U(1)$ gauge 
fields in this theory couples to a linear combination of the $n$ FI 
parameters, which then has the 1-loop renormalization 
\be\label{ra1loop}
r_{\al} = \bigg( \sum_{\al=1}^n \al_a \sum_{i=1}^{n+d} Q_i^a 
\bigg)\cdot \log {\mu\over \Lambda}\ ,
\ee
the $\al_a$ being arbitrary real numbers. The above linear combination 
of the $U(1)$'s is marginal if the coefficient of $r_{\al}$ vanishes, 
\ie, \be
\sum_{\al=1}^n \sum_{i=1}^{n+d} \al_a Q_i^a = 0\ .
\ee
This equation defines a codimension-one hyperplane perpendicular to a 
ray emanating from the origin and passing through the point $(-\sum_i 
Q_i^1, -\sum_i Q_i^2, \ldots, -\sum_i Q_i^n)$ in $r$-space which has 
real dimension $n$. This point lies in the $r_a > 0$ region of 
$r$-space for GLSMs containing only tachyonic operators\footnote{Recall 
that for appropriate isolated nonsupersymmetric orbifolds $\BC^3/\BZ_N\ 
(1,p,q)$, there are no marginal twisted sector states, only tachyons 
(and of course irrelevant operators).}---we henceforth call this the 
$flow$-$ray$.

We can redefine the charge matrix $Q_i^a$ in a useful way that will 
help illustrate some of the physics of the orbifold decay in general. 
Define 
\be\label{Qia'}
{Q_i^a}' \equiv \bigg(\sum_i Q_i^1\bigg) Q_i^a - \bigg(\sum_i 
Q_i^a\bigg) Q_i^1\ , \qquad \qquad a\neq 1\ .
\ee
In other words, we have, \eg, ${Q_1^2}' = N (R_1 - 1) q_1^2 - N (R_2 - 
1) q_1^1$. With this redefinition, we have\ $\sum_i {Q_i^a}' =  (\sum_i 
Q_i^1) (\sum_i Q_i^a) - (\sum_i Q_i^a) (\sum_i Q_i^1) = 0$, \ for 
$a\neq 1$, \ so that the FI parameters coupling to these redefined 
$n-1$ gauge fields have vanishing 1-loop running. Thus there is a 
single relevant direction (along the flow-ray) and $n-1$ marginal 
directions in $r$-space.

Since the FI parameters $r_a$ necessarily are renormalized, we expect
from the structure of the RG flowlines in $r$-space that some phases
will represent $stable$ fixed points while others will be $unstable$
(in the sense of the renormalization group) and will flow to the
stable ones. Thus we expect that the geometry, in a given initial
phase, is in general forced to dynamically evolve to another, more
stable, phase, possibly passing through intermediate phases: this
potential instability and the decay to a stable endpoint geometry is
the spacetime reflection of the renormalization group running of the
FI parameters in the GLSM. By studying various linear combinations
(\ref{ra1loop}), we see that the 1-loop renormalization group flows
drive the system along the single relevant direction to the phases in 
the large $r$ regions of $r$-space, \ie, $r_a\gg 0$, that are adjacent 
to the flow-ray \ $f\equiv (-\sum_i Q_i^1, -\sum_i Q_i^2, \ldots, 
-\sum_i Q_i^n)$, \ or contain it in their interior.

The spacetime reflection of the single relevant (tachyonic) direction
in $r$-space is the existence of a dominant unstable direction for
the decay of the orbifold. The $n-1$ marginal directions of the GLSM
perpendicular to the flow-ray indicate the presence of $n-1$ flat
directions (moduli) of the tachyon potential in spacetime. These flat
directions are in general appropriate combinations of the subsequent
tachyonic twisted states besides the dominant (most relevant)
tachyon. In the example $\BC^3/\BZ_{13}\ (1,2,5)$ that we discuss in
detail in Sec.~3, there are three twisted sector tachyons: the
combinatorics of $Q_i^a$ determines that the two residual flat
directions are in fact the two subsequent twisted states themselves,
besides the most relevant tachyon $T_1$, \ie, the subsequent tachyons
get renormalized under the RG flow induced by $T_1$ and become
marginal.

Note that even if all $r_a$ are initially large, some of them may
become small during RG flow, rendering our semiclassical analysis
untrustworthy.  However, for inital values of $r_a$ whose components
in the marginal directions lie far from the center of the marginal
$(n-1)$-plane, the values of $r_a$ remain large throughout the
RG flow.

The combinatorial geometry of the flow-ray in $r$-space determines the 
physical structure of the tachyon potential. In particular, the 
combinatorics of the charge matrix $Q_i^a$ determines whether the 
flow-ray is in the interior of all possible convex hulls of $n$-tuples 
of vectors in $r$-space or an edge to one or more of them. For 
instance, consider a 2-tachyon system (see figure~\ref{figcone}) where 
a subsequent tachyon, say $T_2$, is marginal, \ie, 
\be
R_2' = R_2\ \bigg(1 + {q_1^2\over q_1^1}{1-R_1\over R_2} \bigg) = 1\ ,
\ee
using the results of \cite{drmknmrp} for the renormalized R-charges of 
subsequent tachyons. This can be simplified and used to show that\ \ 
$q_1^2 \sum_i Q_i^1 = q_1^1 \sum_i Q_i^2$, \ which therefore implies 
that the flow-ray $f\equiv (-\sum_i Q_i^1,-\sum_i Q_i^2) \propto 
(q_1^1,q_1^2)\equiv \phi_1$,\ using (\ref{Qia2gen}), (\ref{Qiagen}). 
This shows that the flow-ray $f$ is the phase boundary $\phi_1$ and 
therefore an edge to two or more convex hulls, \ie, it is not in the 
interior of all convex hulls. Therefore we conclude that the flow-ray 
is the phase boundary between two distinct phases that are connected by 
the single marginal direction perpendicular to the flow-ray $f$. We can 
get additional insight into the structure of these phases: for instance 
if ${q_2^2\over q_2^1} > {q_1^2\over q_1^1}={\sum_i Q_i^2\over \sum_i 
Q_i^1}$, then we have $q_2^2\sum_i Q_i^1 - q_2^1\sum_i Q_i^2 > 0$, 
giving the direction of the flip transition of the previous 
subsection. Note that a flip region such as the one described 
in the previous subsection can always be embedded in a 2-tachyon 
subsystem of the $n$-tachyon system since the flip transition 
always involves flipping only one ``wall'' in the toric fan.
Such an embedding provides useful information about the transition
(although the $2$-tachyon subsystem may not completely decouple
from the other tachyons, so this information must be used with care.)

We will now describe quantum corrections to the classical $r$-space. 
The discussion here is applicable away from the region in $r$-space 
where not all of the $r_a$ are small (\ie, for $r$-values far from the 
origin in $r$-space). The 1-loop renormalization of the FI parameters 
can be expressed \cite{wittenphases, wittenIAS, 
morrisonplesserInstantons} as a $\sigma$-dependent shift in terms of a 
perturbatively quantum-corrected twisted chiral superpotential for the 
$\Sigma_a$ 
\be\label{twistedW}
{\tilde W}(\Sigma) = {1\over 2\sqrt{2}} \sum_{a=1}^n \Sigma_a \bigg( 
i{\hat \tau}_a - {1\over 2\pi} \sum_{i=1}^{d+n} Q_i^a \log (\sqrt{2} 
\sum_{b=1}^n Q_i^b \Sigma_b/\Lambda) \bigg)\ .
\ee
This expression has been obtained by considering the large-$\sigma$ 
region in field space and integrating out those scalars $\psi_i$ that 
are massive here (and their expectation values vanish energetically). 
This leads to a modification in the potential energy \be
U(\sigma) = {e^2\over 2} \sum_{a=1}^n \bigg| i{\hat \tau}_a - 
{\sum_{i=1}^{d+n} Q_i^a \over 2\pi} (\log (\sqrt{2} \sum_{b=1}^n Q_i^b 
\sigma_b/\Lambda) + 1) \bigg|^2\ .
\ee
The singularities predicted classically at the locations of the phase 
boundary hyperplanes arise from the existence of low-energy states at 
large $\sigma$. Now from above, we see that along the single relevant 
direction where $\sum_i Q_i^1\neq 0$, the potential energy has a $|\log 
(\sigma_1)|^2$ growth so that the field space accessible to very 
low-lying states is effectively compact and there is no singularity 
along the single relevant direction given by the flow-ray: in other 
words, the RG flow is smooth along the tachyonic directions for all 
values of $\tau_1$. Thus a phase boundary hyperplane that was a 
singularity classically is simply a label for the boundary of the 
adjacent phases quantum mechanically. Continuing past this non-existent 
``singularity'' to a region where $r_1$ differs in sign from $\sum_i 
Q_i$, we find $\sigma$-vacua at large $|r_1|$. These nonclassical 
Coulomb branch vacua are also possible IR endpoints of the RG flows of 
the GLSM along with the geometric endpoints on the Higgs branch: thus 
tachyon condensation drives an unstable orbifold generically to a 
combination of the geometric Higgs branch vacua and nonclassical 
Coulomb branch vacua. The Higgs branch vacua have a transparent 
interpretation in terms of the partial resolutions of the orbifold 
geometry and are thus describable in principle in terms of the 
nonlinear sigma models describing the orbifold. The nonclassical 
Coulomb branch vacua, crucial in ensuring that quasi-topological 
quantities such as the number of vacua are preserved along the RG flow 
\cite{martinecmoore}, should also in principle be visible from the 
nonlinear sigma model point of view: it would be interesting to 
investigate this question more directly.

On the other hand, along the $n-1$ marginal directions, we have $\sum_i 
{Q_i^b}' = 0,\ b\neq 1$: then there are simply constant shifts 
\be
{\tau^{eff}_b}' = {\tau_b}' + {i\over 2\pi} \sum_{i=1}^n {Q_i^b}' \log 
({Q_i^b}')\ ,
\ee
which is $\sigma$-independent (note that this is valid only in the 
vicinity of one $r_i$ small and all other $r_a$ large). There is a 
singularity for ${\tau^{eff}_b}' = 0$, which exists for a particular 
value of $r_b$ and $\theta_b$, in other words for complex codimension 
one. Thus the distinct phases can be continuously connected by paths in 
$r$-space lying in the marginal hyperplane since we have the freedom of 
turning on a nonzero $\theta$-angle distinct from the singular value.

The extrema of the quantum-corrected superpotential (\ref{twistedW}) 
are given by 
\be\label{sigmaV}
\prod_{i=1}^{d+n}\ \bigg( {\sqrt{2} e\over \Lambda} \sum_{b=1}^n Q_i^b 
\sigma_b \bigg)^{Q_i^a} = e^{2\pi i \tau_a} \equiv q_a\ , \qquad 
a=1,\ldots, n\ .
\ee
These inhomogeneous equations for the $\sigma_a$ can be recast using 
the redefined charges ${Q_i^a}'$ given in (\ref{Qia'}) and redefining 
the $\sigma_a$ in terms of appropriate linear combinations to obtain 
one inhomogeneous equation for the single relevant $U(1)$ (which is the 
corresponding equation for $a=1$ above) and $n-1$ homogeneous equations 
for the marginal $U(1)$'s \be
\prod_{i=1}^{d+n}\ \bigg( \sum_{b=2}^n {Q_i^b}' \sigma_b' 
\bigg)^{{Q_i^a}'} = e^{2\pi i \tau_a'} \equiv q_a'\ , \qquad 
a=2,\ldots, n\ .
\ee
Notice that the cutoff $\Lambda$, which is the signature of the RG 
flow, does not appear in these equations as expected, since these 
contain only marginal directions. These equations contain nontrivial 
information about the nonclassical large-$\sigma$ Coulomb branch vacua 
that exist at the infrared of the RG flow, at large $r_a$.

\section{Phases of a GLSM: $\BC^3/\BZ_{13}\ (1,2,5)$}

In what follows, we will flesh out the above phenomena in detail in a 
specific example: we will describe the Type 0 string theory on the 
unstable nonsupersymmetric orbifold $\BC^3/\BZ_{13}\ (1,2,5)$ studied 
in \cite{drmknmrp} via toric methods (and reviewed in Sec.~1). 
Appropriate GSO projections can be imposed without changing the 
essential points of our discussion here. We restrict attention to the 
three tachyons arising in the $(c_X,c_Y,c_Z)$ ring of twisted states of 
the orbifold: this is equivalent to performing a topological twist that 
retains only the chiral $(c_X,c_Y,c_Z)$ ring, projecting out the 
remainder.

\subsection{The 2-parameter system: two tachyons}

{\bf \emph{The classical phase diagram}}

In this subsection (see also Sec.~2.2), we will ignore the tachyon
$T_3$ for simplicity and focus on the physics of the condensation of
tachyons $T_1$ and $T_8$: this gives a GLSM with gauge group
$U(1)\times U(1)$ whose physics is consistent with the full
three-tachyon analysis of the $U(1)^3$ GLSM we do in a later
subsection. The tachyons $T_1$ and $T_8$ have R-charges $R_1\equiv
({1\over 13},{2\over 13},{5\over 13})={8\over 13}$ and $R_8\equiv
({8\over 13},{3\over 13},{1\over 13})={12\over 13}$.  The vacuum
structure of the theory is governed by two independent FI parameters
coupling to the two $U(1)$'s: these parameters run under the RG flow
induced by the two tachyonic operators in the GLSM. The action of the
$U(1)\times U(1)$ on the fields
$\Psi_i\equiv \phi_1, \phi_2, \phi_3, T_1, T_8$ is given, as in 
(\ref{Qiagen}), by \ $\Psi_i \ra e^{i Q_i^a\lambda}\ \Psi_i$,\ where 
the charge matrix is \be \label{Qia2param}
Q_i^a = \left( \bA{ccccc} 1 & 2 & 5 & -13 & 0  \\ 8 & 3 & 1 & 0 & -13  
\\ \eA \right) .
\ee
(We shall use this version of the charge matrix which makes the tachyons
visible, but a more accurate version would replace the first row by
\be \label{newrow} \left(\bA{ccccc} 3&1&0&1&-5 \eA \right), \ee
which is $-\frac1{13}$ times the first row plus $\frac5{13}$ times the 
second.)

The classical vacua of the theory, \ie, the phase diagram of the 
corresponding GLSM, are found by studying the D-terms 
\bea\label{U1xU1Dterms}
D_1 &=& |\phi_1|^2 + 2 |\phi_2|^2 + 5 |\phi_3|^2 - 13 |T_1|^2 - r_1 = 0 
,
\nonumber\\
D_2 &=& 8 |\phi_1|^2 + 3 |\phi_2|^2 + |\phi_3|^2 - 13 |T_8|^2 - r_2 = 0 
.
\eea
In the region $r_1, r_2 \ll 0$, nonzero vacuum expectation values are 
forced upon both $T_1$ and $T_8$: these break the $U(1)\times U(1)$ 
down to a discrete group $\BZ_{13}$ with generator $\omega=e^{2\pi 
i/13}$, whose action on the coordinate fields $\phi_i$ is given by \ $ 
g: (\phi_1,\phi_2,\phi_3) \ra (\omega \phi_1, \omega^2 \phi_2, \omega^5 
\phi_3)$. This is thus the unresolved $\BC^3/\BZ_{13}\ (1,2,5)$ 
orbifold phase. In this phase, the point $\{T_1=0\ \cup\ T_8=0\}$ must 
be excluded from the moduli space since there is no solution to the 
D-term equations at that point: this point is the excluded set in this 
phase.

Now the various other phases can be obtained by focussing on different 
linear combinations of the two $U(1)$'s and realizing the D-terms for 
them, and thereby the phases. The linear combinations 
\bea\label{D'2param}
D_1' &=& {1\over 13} (8D_1 - D_2) = |\phi_2|^2 + 3 |\phi_3|^2 - 8 
|T_1|^2 + |T_8|^2 - {8 r_1 - r_2\over 13} = 0 ,\nonumber\\
D_2' &=& {1\over 13} (3D_1 - 2D_2) = - |\phi_1|^2 + |\phi_3|^2 - 3 
|T_1|^2 + 2 |T_8|^2 - {3 r_1 - 2 r_2\over 13} = 0 ,\\
D_3' &=& {1\over 13} (- D_1 + 5D_2) = 3 |\phi_1|^2 + |\phi_2|^2 + 
|T_1|^2 - 5 |T_8|^2 + {r_1 - 5 r_2\over 13} = 0 ,\nonumber
\eea
show that the rays drawn from the origin $(0,0)$ out to $(1,8), (2,3), 
(5,1)$ are phase boundaries, alongwith the rays $(-13,0), (0,-13)$ 
gleaned from (\ref{U1xU1Dterms}).

\begin{figure}
\bc
\epsfig{file=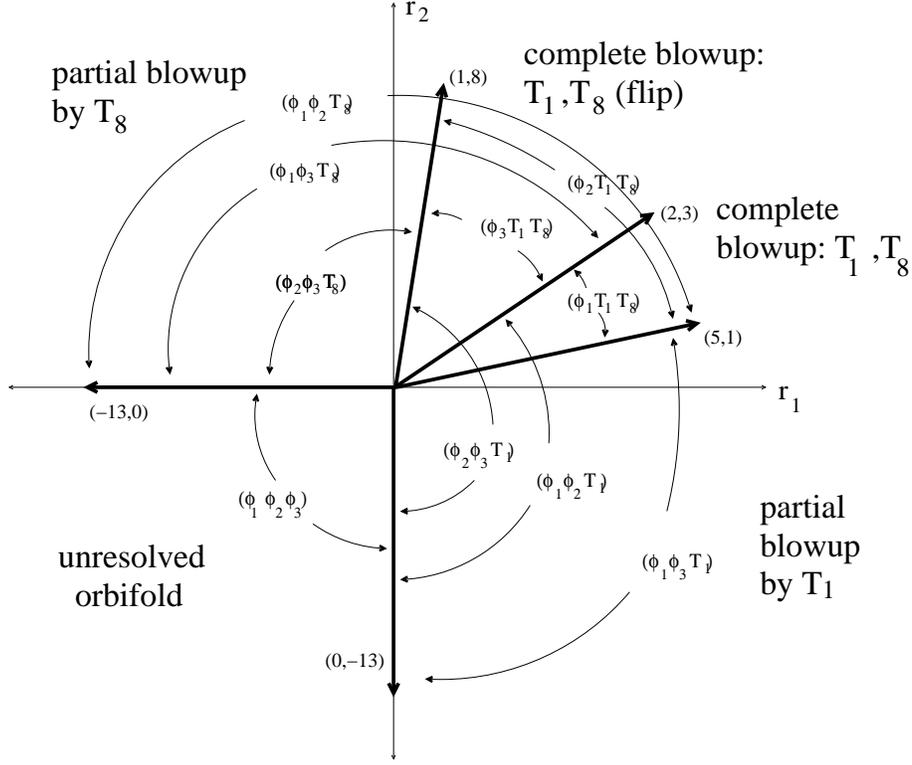, width=12cm}
\caption{Phases of $\BC^3/\BZ_{13}\ (1,2,5)$: the 2-parameter system of 
the two tachyons $T_1, T_8$. Also shown are the various coordinate 
charts characterizing each of the five phases.}
\label{fig2}
\ec
\end{figure}

To illustrate how the moduli space of the GLSM realizes these phases, 
let us study, for instance, the convex hull defined by the rays $(2,3)$ 
and $(5,1)$ (see Figure~\ref{fig2}): then we have $3r_1-2r_2 \gg 0,\ 
-r_1+5r_2 \gg 0$,\ \ie, ${1\over 5}r_1 < r_2 < {3\over 2}r_1$, which 
automatically implies $r_1,\ r_2,\ -r_1 + 8 r_2 \gg 0$. Then solutions 
to $D_2'$ and $D_3'$ exist if at least one of $\{ \phi_3, T_8\}$ and 
$\{ \phi_1, \phi_2, T_1\}$ have nonzero expectation values. Consider 
the region of moduli space where $\phi_3$ and $\phi_2$ alone acquire 
expectation values: then at low energies, the remaining massless fields 
are $\phi_1, T_1, T_8$ which yield a description of the coordinate 
chart $(\phi_1, T_1, T_8)$. Note that this is compatible with the 
constraints given by $D_1, D_2, D_1'$. Likewise, the region of moduli 
space where $\phi_3$ and $\phi_1$ alone acquire expectation values 
leaves the remaining massless fields describing the coordinate chart 
$(\phi_2, T_1, T_8)$. Similarly, the regions where $\phi_3, T_1$ or 
$T_8, \phi_1$ or $T_8, \phi_2$ alone acquire expectation values yield 
descriptions of spaces with coordinate charts $(\phi_1, \phi_2, T_8)$, 
$(\phi_2, \phi_3, T_1)$ and $(\phi_1, \phi_3, T_1)$, 
respectively\footnote{Following this logic would lead one to suspect 
the existence of a region in moduli space for this phase in $r$-space 
where $T_1, T_8$ alone acquire expectation values: however from $D_1', 
D_2', D_3'$, we obtain conditions $|T_8|^2 \gg 8 |T_1|^2, \ |T_8|^2 \gg 
{3\over 2} |T_1|^2, \ |T_1|^2 \gg 5 |T_8|^2$, which do not admit any 
solution thus precluding the existence of the corresponding chart.}. If 
multiple fields acquire nonzero expectation values, the remaining 
massless fields describe overlaps of some of these coordinate charts 
obtained above. Thus the convex hull defined by rays $(2,3)$ and 
$(5,1)$ admits a description of the spacetime geometry by the 
collection of coordinate charts $\{\ (\phi_1, T_1, T_8),\ (\phi_2, T_1, 
T_8),\ (\phi_1, \phi_2, T_8),\ (\phi_2, \phi_3, T_1),\ (\phi_1, \phi_3, 
T_1)\ \}$. From Figure~\ref{fig1}, we see that this collection of 
charts describes the complete resolution of the orbifold obtained by 
the condensation of tachyons $T_1$ and $T_8$ in that sequence: recall 
that this is the most relevant tachyon sequence which leads to the 
least singular residual spacetime geometry.

By a similar analysis of the D-term equations, we can get a physical 
understanding of the geometry described in each of the other phases of 
the GLSM.

A simple operational method\footnote{We mention here for convenience 
that a 2-dim convex hull is defined as the interior of a region bounded 
by two rays emanating out from the origin such that the angle subtended 
by them is less than $\pi$.} to realize the results of the above 
analysis of the D-terms for the phase boundaries and the phases of the 
GLSM is the following: read off each column from the charge matrix 
$Q_i^a$ (eqn.(\ref{Qia2param})) as a ray emanating outwards from the 
origin $(0,0)$ in $(r_1,r_2)$-space, representing a phase boundary. 
Then the various phases in this 2-parameter system are simply given by 
the convex hulls bounded by any two of the (in this case, five) phase 
boundaries represented by the rays $\phi_1\equiv (1,8),\ \phi_2\equiv 
(2,3),\ \phi_3\equiv (5,1),\ T_1\equiv (-13,0),\ T_8\equiv (0,-13)$. 
These phase boundaries divide the $(r_1,r_2)$-plane into five phase 
regions. There are several possible overlapping coordinate charts 
characterizing any given phase of these five described as a convex hull 
of two phase boundaries: these can be read off by noting all the 
possible convex hulls that contain the convex hull (phase) in question.

For instance, consider the convex hull $\{\phi_2, \phi_3\}$: this is 
also contained in the convex hulls $\{\phi_1, \phi_3\},\ \{T_1, 
\phi_3\},
\ \{\phi_2, T_8\}$ and $\{\phi_1, T_8\}$. The coordinate chart
describing a particular convex hull, say $\{\phi_2,\phi_3\}$ is read
off as the complementary set $\{\phi_1,T_1,T_8\}$. Thus the full set
of coordinate charts characterizing the toric variety in the phase
given by the convex hull $\{\phi_2, \phi_3\}$ is \ $\{\ 
(\phi_1,T_1,T_8),\ (\phi_1,\phi_3,T_1),\ (\phi_2,T_1,T_8),\
(\phi_1,\phi_2,T_8),\ (\phi_2,\phi_3,T_1)\ \}$.  From
figure~\ref{fig1}, it is clear that this collection of coordinate
charts describes the toric variety that is obtained by the complete
resolution of the orbifold obtained by turning on the blowup modes 
corresponding to the twist fields $T_1$ and $T_8$ (in that order) in 
the orbifold conformal field theory. As mentioned earlier, this is the 
most relevant tachyon sequence resulting in the least singular 
spacetime geometry.

Likewise we can analyze the other phases of the theory. Consider the 
convex hull $\{\phi_1, \phi_2\}$: this is also contained in the convex 
hulls $\{\phi_1, \phi_3\},\ \{\phi_1, T_8\},\ \{T_1, \phi_2\},\ \{T_1, 
\phi_3\}$. Thus the collection of coordinate charts describing the 
toric variety in this phase is \[\{\ (\phi_3,T_1,T_8),\ 
(\phi_2,T_1,T_8),\  (\phi_2,\phi_3,T_1),\  (\phi_1,\phi_3,T_8),\ 
(\phi_1,\phi_2,T_8)\ \}.\]
{}From figure~\ref{fig1}, we see that this
is the description of the complete resolution of the orbifold when the 
tachyon $T_8$ condenses first, followed by $T_1$: \ie, we subdivide the 
toric fan in the sequence $T_8, T_1$, corresponding to the flip.

The convex hull $\{T_8, \phi_3\}$ is contained in the convex hulls 
$\{T_8, \phi_1\},\ \{T_8, \phi_2\}$, giving the collection of 
coordinate charts as $\{\ (\phi_1,\phi_2,T_1),\ (\phi_2,\phi_3,T_1),\ 
(\phi_1,\phi_3,T_1)\ \}$: this, from figure~\ref{fig1}, corresponds to 
the partial blowup by $T_1$, \ie, this is the endpoint of condensation 
of the tachyon $T_1$ alone.

Similarly, the convex hull $\{T_1, \phi_1\}$ is contained in the convex 
hulls $\{T_1, \phi_2\},\ \{T_1, \phi_3\}$, giving the collection of 
coordinate charts  $\{\ (\phi_2,\phi_3,T_8),\ (\phi_1,\phi_3,T_8),\ 
(\phi_1,\phi_2,T_8)\ \}$: this, from figure~\ref{fig1}, corresponds to 
the partial blowup by $T_8$, \ie, this is the endpoint of condensation 
of the tachyon $T_8$ alone.

Finally the convex hull $\{T_1, T_8\}$ is not contained in any other 
convex hull: it gives the chart $\{\ (\phi_1,\phi_2,\phi_3)\ \}$ that 
describes the unresolved orbifold singularity.\\

\noindent
{\bf \emph{Flowlines}}

The two tachyons $T_1$ and $T_8$ have R-charges $R_1={8\over 13}$ and 
$R_1={12\over 13}$. Then the 1-loop running of the two FI parameters in 
this case (see (\ref{r1r21loop})) is given by \be
r_1(\mu) = -5 \cdot \log {\mu \over \Lambda}\ , \qquad \qquad r_2(\mu) 
= -1 \cdot \log {\mu \over \Lambda}\ .
\ee
Now we can choose various bases for the $U(1)\times U(1)$ gauge fields 
by choosing appropriate linear combinations: these will therefore 
couple to the corresponding linear combinations of the FI parameters in 
the GLSM. The generic linear combination of the two FI parameters 
$\al_1 r_1 + \al_2 r_2$ has the 1-loop running \be\label{flowray51}
\al_1 r_1 + \al_2 r_2 = -(5 \al_1 + \al_2) \cdot \log {\mu \over 
\Lambda}\ .
\ee
{}From the coefficient, we see that this parameter is marginal if \
$5 \al_1 + \al_2 = 0$ : this describes a line in $\al$-space, which is 
perpendicular to the ray $(5,1)$ in $r$-space. Thus we can choose the 
basis for the $U(1)\times U(1)$ so that one linear combination couples 
to a (relevant) FI parameter that has nontrivial renormalization along 
the flow, while the other FI parameter is marginal along the flow.
{}From (\ref{flowray51}), we see that this single relevant direction
(perpendicular to the marginal one), lies along the ray $(5,1)$: this 
is the flow-ray.

Using (\ref{flowray51}), we can get qualitative insight into the 
renormalization group trajectories that cross any of the phase 
boundaries: for instance, (\ref{flowray51}) gives \bea
{3 r_1 - 2 r_2\over 13} &=& - 1 \cdot \log {\mu \over \Lambda}\ , 
\nonumber\\
{8 r_1 - r_2\over 13} &=& - 3 \cdot \log {\mu \over \Lambda}\ , \\
r_1 - (5 + \epsilon) r_2 &=& \epsilon \cdot \log {\mu \over \Lambda}\ .
\nonumber
\eea These equations indicate that at low energies $\mu \ll \Lambda$, 
the 1-loop renormalization group flows drive the system towards the 
large $r$ regions in $r$-space, \ie, $r_1, r_2 \gg 0$, that are 
adjacent to the flow-ray $(5,1)$. There are two such regions: the 
convex hulls $\{T_8, \phi_3\}$\ (for $\epsilon > 0$, \ie, ``below'' the 
flow-ray $(5,1)$) \ and $\{\phi_2, \phi_3\}$ \ (for $\epsilon < 0$, 
\ie, ``above'' the flow-ray $(5,1)$),\ for both of which $r_2 \ll 
{3\over 2} r_1 \ll 8 r_1$ holds. In the process of ending up in one of 
these regions, flowlines may cross one or more of the the 
(semi-infinite) phase boundaries passing through $(-13,0), (1,8), 
(2,3)$ and $(0,-13)$. We again remind the reader here that these 
straight flowlines are only valid away from the origin in $r$-space 
where all $r_a$ are small: this is sufficient for our discussion here.

From the previous subsection, we see that these two phases arising as
the endpoints of flowlines correspond to the decay modes of the 
unstable orbifold that represent the partial resolution by the 
condensation of the tachyon $T_1$ alone and the complete resolution by 
the condensation of the tachyon $T_1$ followed by $T_8$ in that 
sequence (which leads to the minimal singularity). It is interesting to 
interpret this result in terms of the toric description of tachyon 
condensation in this system \cite{drmknmrp} reviewed earlier. Recall 
that the most relevant tachyon is $T_1$ : after its condensation, the 
subsequent tachyon $T_8$ becomes marginal so that the blowup mode it 
corresponds to is now a modulus or flat direction, rather than a 
residual instability. This marginality of the residual (erstwhile 
tachyon) $T_8$ reflects the marginal $U(1)$ direction perpendicular to 
the tachyonic flow-ray $(5,1)$ in the GLSM. Thus the renormalization 
group in the GLSM drives the system along the flow-ray, maintaining a 
flat direction perpendicular to it. On the other hand, note that the 
tachyon $T_1$ remains tachyonic after condensation of $T_8$, so that 
the partial resolution by $T_8$ alone cannot result: the subsequent 
resolution by $T_1$ is $not$ a flat direction but a residual 
instability.

This interpretation is further strengthened by the following: a
flowline emanating in the orbifold phase (\ie, the convex hull $\{T_1,
T_8\}$) $below$ the flow-ray $(5,1)$ necessarily crosses the
(semi-infinite) phase boundary $(0,-13)$ to end up in the convex hull
$\{T_8, \phi_3\}$. This trajectory corresponds to the most relevant
tachyon $T_1$ turning on and condensing, resulting in a partial
resolution. On the other hand, a flowline emanating $above$
the flow-ray $(5,1)$ crosses the (semi-infinite) phase boundaries
$(-13,0),\ (1,8)$ and $(2,3)$ to end up in the convex hull $\{\phi_2,
\phi_3\}$.  This trajectory corresponds to the less relevant tachyon
$T_8$ turning on first: this commences the partial resolution of the
orbifold by the subdivision corresponding to $T_8$. As the flowline
crosses the boundary $(1,8)$, the more relevant tachyon $T_1$ turns on
which leads to the further resolution corresponding to the subsequent
subdivision by $T_1$ : this is the phase represented by the convex
hull $\{\phi_1, \phi_2\}$ corresponding to the flipped subdivision
($T_8$ followed by $T_1$). However now the flowline crosses the
boundary $(2,3)$ to end up finally in the phase $\{\phi_2, \phi_3\}$ :
note that this now represents the complete resolution by the tachyons
but in the sequence $T_1$ followed by $T_8$. Thus in crossing the
phase boundary $(2,3)$, there occurs a flip transition in which the
sequence of partial resolutions of the orbifold by the two tachyons
reverses, from $T_8$ followed by $T_1$ to $T_1$ and then $T_8$.
Correspondingly the degree of singularity also changes to the less
singular resolution.

We note here that the discussion above is just a detailed description 
in this example of the generalities of the previous section.

\subsection{Including all three tachyons: the 3-parameter system}

In this section, we study the decay of the $\BC^3/\BZ_{13}\ (1,2,5)$ 
orbifold including all three tachyons present, \ie, $T_1, T_8, T_3$ : 
these have R-charges $R_1\equiv ({1\over 13},{2\over 13},{5\over 
13})={8\over 13}$ ,\ $R_8\equiv ({8\over 13},{3\over 13},{1\over 
13})={12\over 13}$ and $R_3\equiv ({3\over 13},{6\over 13},{2\over 
13})={11\over 13}$. The GLSM in question has gauge group $U(1)^3$. The 
vacuum structure of the theory is governed by three independent FI 
parameters coupling to the three $U(1)$'s: these parameters run under 
the RG flow induced by the three tachyonic operators in the GLSM. The 
action of the $U(1)^3$ on the fields $\Psi_i\equiv \phi_1, \phi_2, 
\phi_3, T_1, T_8, T_3$ is given, as in (\ref{Qiagen}), by \ $\Psi_i \ra 
e^{i Q_i^a\lambda}\ \Psi_i$ ,\ where the charge matrix is \be 
\label{Qia3param}
Q_i^a = \left( \bA{cccccc} 1 & 2 & 5 & -13 & 0 & 0 \\ 8 & 3 & 1 & 0 & 
-13 & 0 \\ 3 & 6 & 2 & 0 & 0 & -13 \\ \eA \right) .
\ee
(Again, we shall use this form due to the visibility of the tachyons, 
but
in the more accurate form the first row would be replaced by 
(\ref{newrow})
and the third row would be replaced by
\be \label{newrow2} \left(\bA{cccccc} 1&0&0&0&-2&1 \eA \right), \ee
which is $\frac2{13}$ times the second row plus $-\frac1{13}$ times the 
second.)

The D-term equations in this case are \bea\label{U1xU1xU1Dterms}
D_1 &=& |\phi_1|^2 + 2 |\phi_2|^2 + 5 |\phi_3|^2 - 13 |T_1|^2 - r_1 = 0 
,
\nonumber\\
D_2 &=& 8 |\phi_1|^2 + 3 |\phi_2|^2 + |\phi_3|^2 - 13 |T_8|^2 - r_2 = 0 
,\\
D_3 &=& 3 |\phi_1|^2 + 6 |\phi_2|^2 + 2 |\phi_3|^2 - 13 |T_3|^2 - r_3 = 
0 .
\nonumber
\eea
Similarly we can obtain the counterparts of (\ref{D'2param}) of the 
2-parameter analysis by studying the D-term equations obtained for 
various linear combinations of the three $U(1)$'s (the subscripts 
denote the field(s) eliminated choosing two of the three $D_i$s above, 
which are denoted by the superscripts) \bea\label{D'3param}
{D_2^{12}}' &=& -|\phi_1|^2 + |\phi_3|^2 - 3 |T_1|^2 + 2 |T_8|^2 - {3 
r_1 - 2 r_2\over 13} = 0\ ,\nonumber\\
{D_{2,3}^{23}}' &=& |\phi_1|^2 - 2 |T_8|^2 - |T_3|^2 - {2 r_2 - 
r_3\over 13} = 0\ ,\nonumber\\
{D_{2,1}^{13}}' &=& |\phi_3|^2 - 3 |T_1|^2 + |T_3|^2 - {3 r_1 - 
r_3\over 13} = 0\ ,\nonumber\\
{D_1^{12}}' &=& |\phi_2|^2 + 3 |\phi_3|^2 - 8 |T_1|^2 + |T_8|^2 - {8 
r_1 - r_2\over 13} = 0\ ,\\
{D_1^{23}}' &=& -3 |\phi_2|^2 - |\phi_3|^2 - 3 |T_8|^2 + 8 |T_3|^2 - {3 
r_2 - 8 r_3\over 13} = 0\ ,\nonumber\\
{D_3^{12}}' &=& -3 |\phi_1|^2 - |\phi_2|^2 - |T_1|^2 + 5 |T_8|^2 - {r_1 
- 5 r_2\over 13} = 0\ ,\nonumber\\
{D_3^{13}}' &=& -|\phi_1|^2 - 2 |\phi_2|^2 - 2 |T_1|^2 + 5 |T_3|^2 - {2 
r_1 - 5 r_3\over 13} = 0\ .\nonumber
\eea
There are potentially ${6\choose 4}=15$ such D-term equations but some 
of them coincide due to the fact that some of the rays are coplanar, 
giving ten independent equations in (\ref{U1xU1xU1Dterms}), 
(\ref{D'3param}) above.

\begin{figure}
\bc
\epsfig{file=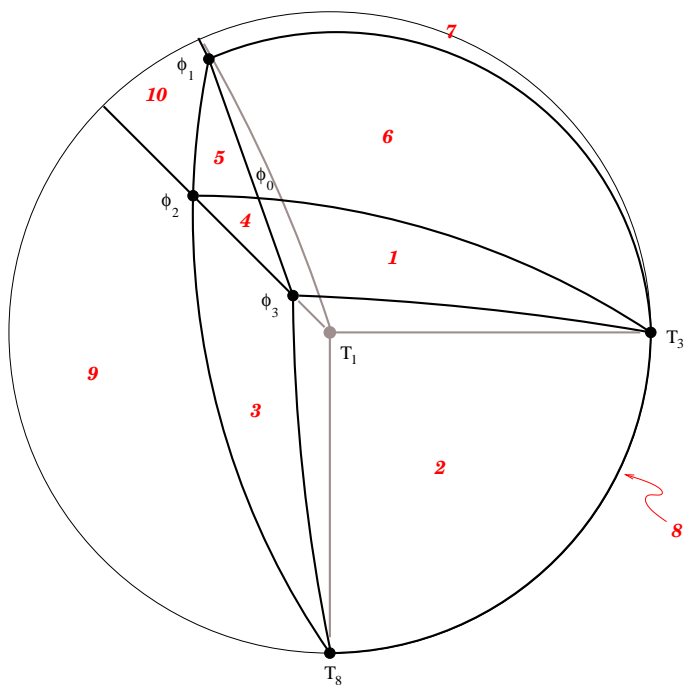, width=9cm}
\caption{Phases of $\BC^3/\BZ_{13}\ (1,2,5)$: the 3-parameter system of 
the three tachyons $T_1, T_8, T_3$. The seven intersection points\ 
$\phi_1,\ \phi_2,\ \phi_3,\ \phi_0,\ T_1,\ T_8,\ T_3$\ of the phase 
boundary hyperplanes are shown as points on a unit sphere around the 
origin in $r$-space, along with the ten phases arising as convex hulls 
bounded by triples of the points.}
\label{fig3}
\ec
\end{figure}

As before, the solutions to these D-term equations give collections of 
coordinate charts that characterize in general distinct toric 
varieties, depending on the FI parameters $r_1, r_2, r_3$: \ie, varying 
$r_1, r_2$ and $r_3$ realizes several distinct phases of the theory 
(see figure~\ref{fig3}). Analyzing the theory along the same lines as 
for the 2-parameter case, we find several distinct phase boundaries 
obtained as planes spanned by pairs of the phase bounding vectors 
$\phi_1\equiv (1,8,3), \ \phi_2\equiv (2,3,6), \ \phi_3\equiv (5,1,2), 
\ \ T_1\equiv (-13,0,0),\ T_8\equiv (0,-13,0),\ T_3\equiv (0,0,-13)$: 
these can be read off as the six columns in the charge matrix 
(\ref{Qia3param}). These planes defining phase boundaries intersect in 
pairs along rays emanating from the origin in $r$-space. The set of 
these intersection rays includes the six vectors $\phi_i, T_j$ above 
but also contains a new ray emanating from the origin out to the vector 
$\phi_0\equiv (6, 9, 5)$, which arises as the intersection of the 
planes $\{ \phi_1, \phi_3\}$ and $\{ \phi_2, T_3 \}$. The geometric 
relations amongst these vectors can be gleaned effectively by treating 
them as toric data. In particular, we see that certain subsets of these 
vectors are coplanar: specifically $\{ \phi_1, \phi_2, T_8 \}$ are 
coplanar, as are $\{ \phi_2, \phi_3, T_1 \}$, in other words, \ ${\rm 
det} (\phi_1, \phi_2, T_8) = 0$, \ and \ ${\rm det} (\phi_2, \phi_3, 
T_1) = 0$ (we can also see this by noting,
\eg, $2\phi_1 - \phi_2 + T_8 = 0$). By writing $\phi_2$ and $\phi_1$ as 
appropriate linear combinations of the other vectors, we find that 
$\phi_2$ lies in the interior of the convex hulls $\{ \phi_1, T_8, T_3 
\}, \ \{ \phi_2, T_8, T_3 \}$ (\eg, we have $\phi_3 = 5 \phi_1 + 3 T_8 
+ T_3$)\ and $\phi_1 \in \{ \phi_3, T_1, T_3 \}$.

The various phases of the GLSM for this 3-parameter system can be 
obtained as in the 2-parameter case, by an analysis of the D-term 
equations, or equivalently by the operational method described in that 
subsection. Each phase in this case is a 3-dimensional convex hull, or 
3-simplex, defined by triples of the set of intersection rays $\{ 
\phi_1, \phi_2, \phi_3, \phi_0, T_1, T_8, T_3 \}$, bounded by the 
intersection planes mentioned above. Figure~\ref{fig3} shows the 
3-dimensional phase diagram which has been drawn by plotting the 
projections of these seven rays on a unit sphere in $r$-space: in other 
words, we have shown the points ${\phi_i\over \| \phi_i \|}, \ 
{T_j\over \| T_j\| }$ as points on the unit sphere\footnote{For the 
benefit of the reader, we mention here that this is probably the most 
convenient method to accurately obtain a visualization of the relations 
between the rays and thence the various phases: figure~\ref{fig3} was 
obtained by first plotting these points on an actual racquetball(!) 
after finding that the stereographic projection onto a 2-plane 
intersecting the unit sphere yielded results
that were difficult to interpret.}. The view depicted in 
figure~\ref{fig3} is obtained by looking head-on ``into'' the sphere 
from the point antipodal to $T_1$, in other words, the line of sight is 
along the line\ $(1,0,0)\ra (-1,0,0)$, \ in the direction of the ray 
$T_1$. The dark solid arcs are the intersections of the phase boundary 
hyperplanes with the unit sphere that lie on the hemisphere in direct 
view (\ie, containing the point $\phi_3\equiv (5,1,2)$) while the 
lighter arcs are the intersections that lie on the antipodally opposite 
hemisphere containing the point $T_1\equiv (-13,0,0)$.

We now outline the operational method described in the 2-parameter 
discussion to obtain the ten phases in Figure~\ref{fig3}. Consider, for 
instance, the phase $r_1, r_2, r_3 \ll 0$, \ie, the convex hull $\{ 
T_1, T_8, T_3 \}$: these fields alone acquire nonzero expectation 
values and the remaining massless fields describe the coordinate chart 
$(\phi_1, \phi_2, \phi_3)$, \ie, the unresolved orbifold singularity. 
Likewise, in the convex hull $\{ \phi_1, \phi_2, T_1 \}$, the residual 
massless fields describe the chart $(\phi_3, T_8, T_3)$. The procedure 
is clear: each convex hull $\{ \psi_1, \psi_2, \psi_3 \}$ yields a 
description of a coordinate chart given by the complement of these 
fields in the collection of the six vectors $\{ \phi_i, T_j \}, \ 
i=1,2,3, \ j=1,8,3$. In addition to this of course, we note as before
that a given convex hull may be itself contained in other convex
hulls: thus each convex hull yields a collection of coordinate charts
which then describes a toric variety. Furthermore, the convex hulls
thus obtained using triples of these six vectors in general intersect
giving possibly new vectors: as we have seen above, the only new
vector in this case is $\phi_0$. By carefully studying the 
3-dimensional phase diagram in Figure~\ref{fig3}, we can see that the 
ten distinct phases\footnote{Note that Figure~\ref{fig3} can be thought 
of as a picture of a triangulation of the unit sphere where each 
triangle is an intersection of the corresponding convex hull with the 
unit sphere. Then we know the Euler characteristic for the sphere to be 
$\chi=2=Faces+Vertices-Edges$, which gives the $10+7-2=15$ edges or 
phase boundaries shown in Figure~\ref{fig3}.} are as given in 
Table~\ref{table3paramphases1} and Table~\ref{table3paramphases2}: the
first column labels the phase number as shown in figure~\ref{fig3},
the second column gives the phases as intersections of convex hulls,
the third gives the collection of coordinate charts valid in each
phase while the fourth column then gives the interpretation of each
phase in accord with Figure~\ref{fig1}.

\vspace{1mm}
\begin{table}[htb]
\bc
\caption{Stable phases of $\BC^3/\BZ_{13}\ (1,2,5)$}
\label{table3paramphases1}
\begin{tabular}{|c| p{5.5cm} | p{4.5cm} | p{4cm} |}
\hline
1 & $\{ \phi_3, \phi_0, T_3 \} = \{ \phi_1, T_3, \phi_3 \} \cap \{ 
\phi_2, T_3, \phi_3 \} \cap \{ \phi_3, T_1, T_3 \} \cap \{ \phi_1, T_8, 
T_3 \} \cap \{ \phi_2, T_8, T_3 \}$ & $(\phi_2, T_1, T_8), \ (\phi_1, 
T_1, T_8), $ \ $ (\phi_1, \phi_2, T_8), \ (\phi_2, \phi_3, T_1), $ \ 
$(\phi_1, \phi_3, T_1)$ & Partial resolution by $T_1$ followed by $T_8$ 
(minimal singularity) \\ \hline
2 & $\{ \phi_3, T_3, T_8 \} = \{ \phi_3, T_3, T_8 \} \cap \{ \phi_1, 
T_3, T_8 \} \cap \{ \phi_2, T_3, T_8 \}$ & $ (\phi_1, \phi_2, T_1), \ 
(\phi_2, \phi_3, T_1), $ \
  $(\phi_1, \phi_3, T_1)$ & Partial resolution by $T_1$ \\
\hline
3 & $\{ \phi_3, \phi_2, T_8 \} = \{ \phi_2, \phi_3, T_8 \} \cap \{ 
\phi_1, \phi_3, T_8 \} \cap \{ \phi_3, T_1, T_8 \} \cap \{ \phi_1, T_8, 
T_3 \} \cap \{ \phi_2, T_8, T_3 \}$ &
$(\phi_1, T_1, T_3), \ (\phi_2, T_1, T_3), $ \ $ (\phi_1, \phi_2, T_3), 
\ (\phi_2, \phi_3, T_1), $ \ $(\phi_1, \phi_3, T_1)$ & Partial 
resolution by $T_1$ and $T_3$ \\
\hline
4 & $\{ \phi_3, \phi_2, \phi_0 \} = \{ \phi_1, T_8, \phi_3 \} \cap \{ 
\phi_2, T_3, \phi_3 \} \cap \{ \phi_1, T_8, T_3 \} \cap \{ \phi_2, 
\phi_3, T_3 \} \cap \{ \phi_3, T_1, T_3 \} \cap \{ \phi_2, T_3, T_8 \} 
\cap
\{ \phi_1, \phi_2, \phi_3 \}$ & $(\phi_2, T_1, T_3), \ (\phi_1, T_1, 
T_8), $ \ $ (\phi_2, \phi_3, T_1), \ (\phi_1, T_1, T_8), $ \ $(\phi_1, 
\phi_2, T_8), \ (\phi_1, \phi_3, T_1), $ \ $(T_1, T_8, T_3)$ & Complete 
resolution by $T_1$ followed by $T_8, T_3$ (minimal singularity) \\
\hline
\end{tabular}
\ec
\end{table}
\vspace{1mm}
\vspace{1mm}
\begin{table}[htb]
\bc
\caption{Unstable phases of $\BC^3/\BZ_{13}\ (1,2,5)$}
\label{table3paramphases2}
\begin{tabular}{|c| p{5.5cm} | p{4.5cm} | p{4cm} |}
\hline
5 & $\{ \phi_0, \phi_2, \phi_1 \} = \{ \phi_1, \phi_2, \phi_3 \} \cap 
\{ \phi_1, \phi_2, T_3 \} \cap \{ \phi_1, \phi_3, T_1 \} \cap \{ 
\phi_1, T_8, \phi_3 \} \cap \{ \phi_2, T_3, T_1 \} \cap \{ \phi_3, T_3, 
T_1 \} \cap \{ \phi_1, T_1, T_8 \}$ & $(T_1, T_8, T_3), \ (\phi_3, T_1, 
T_8), $ \ $ (\phi_2, T_3, T_8), \ (\phi_2, T_1, T_3), $ \ $(\phi_1, 
\phi_3, T_8), \ (\phi_1, \phi_2, T_8), $ \ $(\phi_2, \phi_3, T_1)$ & 
Complete resolution by $T_8$ followed by $T_1, T_3$ (flip) \\
\hline
6 & $\{ \phi_0, \phi_1, T_3 \} = \{ \phi_1, \phi_2, T_3 \} \cap \{ 
\phi_1, \phi_3, T_3 \} \cap \{ \phi_3, T_1, T_3 \} \cap \{ \phi_2, T_1, 
T_3 \} \cap \{ \phi_1, T_8, T_3 \}$ & $(\phi_3, T_1, T_8), \ (\phi_2, 
T_1, T_8), $ \ $ (\phi_1, \phi_2, T_8), \ (\phi_1, \phi_3, T_8), $ \ 
$(\phi_2, \phi_3, T_1)$ & Partial resolution by $T_8$ followed by $T_1$ 
(flip) \\ \hline
7 & $\{ \phi_1, T_1, T_3 \} = \{ \phi_1, T_1, T_3 \} \cap \{ \phi_2, 
T_1, T_3 \} \cap \{ \phi_3, T_1, T_3 \}$ & $ (\phi_2, \phi_3, T_8), \ 
(\phi_1, \phi_3, T_8), $ \
  $(\phi_1, \phi_2, T_8)$ & Partial resolution by $T_8$ \\
\hline
8 & $\{ T_1, T_8, T_3 \}$ & $(\phi_1, \phi_2, \phi_3)$ & Unresolved 
orbifold \\
\hline
9 & $\{ \phi_2, T_1, T_8 \} = \{ \phi_2, T_1, T_8 \} \cap \{ \phi_3, 
T_1, T_8 \} \cap \{ \phi_1, T_1, T_8 \}$ & $ (\phi_1, \phi_3, T_3), \ 
(\phi_1, \phi_2, T_3), $ \
  $(\phi_2, \phi_3, T_3)$ & Partial resolution by $T_3$ \\
\hline
10 & $\{ \phi_1, \phi_2, T_1 \} = \{ \phi_1, \phi_2, T_1 \} \cap \{ 
\phi_1, T_1, T_8 \} \cap \{ \phi_3, \phi_1, T_1 \} \cap \{ \phi_2, T_1, 
T_3 \} \cap \{ \phi_3, T_1, T_3 \}$ &
$(\phi_3, T_3, T_8), \ (\phi_2, \phi_3, T_3), $ \ $ (\phi_2, T_8, T_3), 
\ (\phi_1, \phi_3, T_8), $ \ $(\phi_1, \phi_2, T_8)$ & Partial 
resolution by $T_8$ and $T_3$ \\
\hline
\end{tabular}
\ec
\end{table}
\vspace{1mm}
The 1-loop running of the FI parameters \be
r_a(\mu) = \biggl(\sum_i Q_i^a\biggr) \cdot \log {\mu \over \Lambda} \ 
, \qquad \qquad a=1,2,3 ,
\ee
is given in this case by \bea
r_1(\mu) &=& -5 \cdot \log {\mu \over \Lambda} \ , \nonumber\\
r_2(\mu) &=& -1 \cdot \log {\mu \over \Lambda} \ , \\
r_3(\mu) &=& -2 \cdot \log {\mu \over \Lambda} \ , \nonumber
\eea
Thus the generic linear combination of the three $U(1)$'s couples to a 
FI parameter \be\label{r1r2r31loop}
\al_1 r_1 + \al_2 r_2 + \al_3 r_3 = -(5 \al_1 + \al_2 + 2 \al_3) \cdot 
\log {\mu \over \Lambda}\ ,
\ee
which is marginal if \ $5 \al_1 + \al_2 + 2 \al_3 = 0$ : this now 
describes a plane perpendicular to the ray emanating from the origin 
out to $(5,1,2)$ in $r$-space. Thus we can choose a basis for the 
charges of the fields under the three $U(1)$'s so that one linear 
combination is relevant, coupling to a FI parameter that has nontrivial 
renormalization along the flow, while the other two FI
parameters are marginal along the flow. This single relevant direction
defining the directionality of all flowlines lies along the ray
emanating from the origin $(0,0,0)$ passing through $(5,1,2)$, in
other words, $\phi_3$ can be identified as the flow-ray in
3-dimensional $r$-space.  By studying various linear combinations
(\ref{r1r2r31loop}), we see that the 1-loop renormalization group
flows drive the system to the large $r$ regions in $r$-space, \ie,
$r_1,r_2,r_3\gg 0$, that are adjacent to the flow-ray $\phi_3\equiv
(5,1,2)$. From Figure~\ref{fig3}, we see that there are four such
phases, labelled by ${\bf 1, 2, 3, 4}$: these are the convex hulls $\{ 
\phi_3, \phi_0, T_3 \}, \{ \phi_3, T_3, T_8 \}, \{ \phi_3, T_8, \phi_2 
\}, \{ \phi_3, \phi_2, \phi_0 \}$,
respectively. These four stable phases are listed in 
Table~\ref{table3paramphases1}.

The other six of the ten represent unstable phases, listed in
Table~\ref{table3paramphases2}: renormalization group flowlines
emanate from these phases and end on the stable ones. Thus for initial
conditions for the geometry as any of these phases, small fluctuations
will force the geometry to dynamically evolve to one of the four
stable phases.

\subsection{The space of geometries and the RG trajectories}

As in the 2-parameter case, it is interesting to revisit the toric 
description of tachyon condensation to shed light on the stability of 
the four phases and instability of the other six. Recall that $T_1$ was 
the most relevant tachyon: thus the dominant decay channel for the 
unstable orbifold involves the condensation of $T_1$ first, in other 
words, the partial resolution of the orbifold by $T_1$. After 
condensation of $T_1$, subsequent tachyons $T_8$ and $T_3$ become 
marginal so that the blowup modes they correspond to are moduli, not 
residual instabilities. Thus subsequent resolutions by $T_8$ or $T_3$ 
represent flat directions of the full tachyon potential: the existence 
of these two independent flat directions in spacetime is reflected by 
the existence of the plane of the two marginal $U(1)$'s in the 
GLSM\footnote{In general, it is possible that the flat directions in 
spacetime are appropriate combinations of the subsequent tachyonic 
twisted states in the GLSM.}. Thus the renormalization group drives the 
system to the large $r$ region along the direction of the flow-ray, 
preserving a plane of flat directions. The possibility of turning on 
the residual moduli yields four possible distinct stable endpoints, 
listed in Table~\ref{table3paramphases1}. It is also of course possible 
for the tachyons to condense in other sequences. In this case however, 
recall that the tachyon $T_1$ remains
tachyonic after condensation of either $T_8$ or $T_3$: in other words,
the tachyon $T_1$, if subsequent, is always a residual instability,
not a flat direction. This is reflected in the fact that all of the
stable phases involve (partial) resolutions by the tachyon $T_1$, in
other words, the (most relevant) tachyon $T_1$ always does condense 
eventually.

Further light is shed on the possible trajectories in the space of 
geometries by studying the projections of these phases onto the plane 
of marginal $U(1)$'s. A basis for the two marginal $U(1)$'s is given by 
the charge matrix\footnote{Equivalently we could use the charge matrix 
corresponding to (\ref{Qia'}).} with entries in the two rows given by \ 
${1\over13}({Q_i^1 - 5 Q_i^2}, \ {Q_i^3 - 2 Q_i^2})$ \ with $Q_i^a$ 
given in (\ref{Qia3param}) :
\be \label{margQia}
Q_i^a = \left( \bA{cccccc} -3 & -1 & 0 & -1 & 5 & 0 \\ -1 & 0 & 0 & 0 & 
2 & -1 \\ \eA \right) .
\ee
\begin{figure}
\bc
\epsfig{file=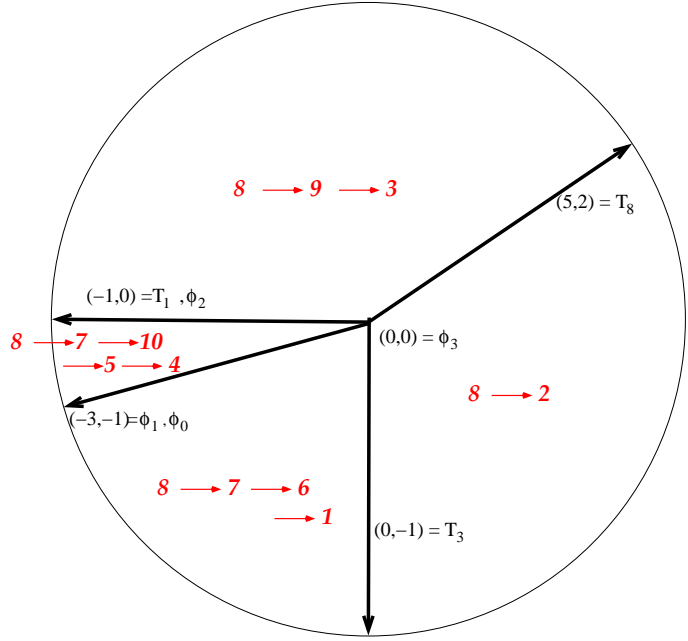, width=9cm}
\caption{The marginal hyperplane and RG trajectories in 
$\BC^3/\BZ_{13}\ (1,2,5)$: shown here is the marginal hyperplane normal 
to the flow-ray $\phi_3$ (taken as the origin). The dark arrows mark 
the intersections of the marginal hyperplane with the phase boundaries. 
Identifying the convex hulls that overlap with each of the regions (see 
figure~\ref{fig3}) gives the four trajectories shown.}
\label{figMargplane}
\ec
\end{figure}
Note that by construction, $\sum_i Q_i^a = 0,\ i=1,2, $ indicating 
marginality. Now treating this as an effective 2-parameter system, we 
can plot the phases in the vicinity of the flow-ray $\phi_3\equiv 
(5,1,2)$, defined as the origin, by drawing rays emerging from the 
origin out to each of the other points (see figure~\ref{figMargplane}). 
Clearly there are four regions. Identifying the subsets of the ten 
phases from Table~\ref{table3paramphases1} and 
Table~\ref{table3paramphases2} that overlap with each of these four 
regions gives the following four trajectories (see 
Figure~\ref{figMargplane}): \begin{enumerate}
\item{Unresolved orbifold [8] $\ra$ partial resolution by $T_8$ [7] 
$\ra$ partial resolution by $T_8$ and $T_1$ (in that sequence) [6] 
$\ra$ partial resolution by $T_1$ and $T_8$ (minimal singularity) [1].} 
\item{Unresolved orbifold [8] $\ra$ partial resolution by $T_1$ [2].} 
\item{Unresolved orbifold [8] $\ra$ partial resolution by $T_3$ [9] 
$\ra$ partial resolution by $T_3$ and $T_1$ [3].} \item{Unresolved 
orbifold [8] $\ra$ partial resolution by $T_8$ [7] $\ra$ partial 
resolution by $T_8$ and $T_3$ [10] $\ra$ partial resolution by $T_8, 
T_3$ and $T_1$ (in that sequence) [5] $\ra$ partial resolution by $T_1, 
T_3$ and $T_8$ (minimal singularity) [4].} \end{enumerate} For the 
3-parameter system, we have seen that there are ten phases or regions 
in $r$-space. From the point of view of the renormalization group for 
the GLSM, each phase is a fixed point, stable or unstable. All of the 
above RG flowlines emanate from the unresolved orbifold phase ${\bf 
8}$, which is the UV fixed point. The infrared fixed point in the 
trajectory $k$ above is the stable phase ${\bf k}$ in 
Table~\ref{table3paramphases1}. The trajectories $1$ and $4$ involve 
flip transitions: these occur in the region given by the 
convex hull $\{ \phi_1, \phi_3, T_8, T_1 \}$ in figure~\ref{fig1}. They 
involve a reversal of the sequence of condensation of tachyons
from $T_8, T_1$ to $T_1, T_8$ and therefore of the blowups
corresponding to them. This involves a blowdown of a divisor and a
blowup of a topologically distinct divisor resulting in a singularity
of smaller degree.

Thus the RG trajectories ${\bf 2}$ and ${\bf 3}$ flow directly from an
unstable UV phase to a stable IR phase. However the trajectories ${\bf
1}$ and ${\bf 4}$ that involve flip transitions may pass arbitrarily
close to a unstable IR fixed point (\ie, an unstable phase) before
eventually flowing to a stable IR fixed point\footnote{This is
reminiscent of the supersymmetric flows that flow from $\BC^3/\BZ_N$
geometries towards $\BC^2/\BZ_N$ before flowing finally to flat space
\cite{knmrp0309}. Along these lines, it would perhaps be interesting to 
ask what a D-brane probe sees during a flip transition.}. In this 
sense, one could think of the blowdown+blowup as the combination of an
irrelevant and a relevant operator in the GLSM. Indeed this is borne
out by the coefficient of the FI parameter $r^{eff}$ for the
corresponding relative $U(1)$ in Sec.~2.2.

The above $\BZ_{13}$ example was in the context of the Type 0 string
theory, where as we have seen, the (nonsupersymmetric) terminal
singularity $\BC^3/\BZ_2\ (1,1,1)$ does appear in the spectrum of
stable phases under tachyon condensation. Similar analyses can be
performed in the context of the Type II string. A key observation here
is that for every stable phase in the Type II theory, the residual
singularities are all supersymmetric. Furthermore, in a stable phase,
some of the sizes of curves and surfaces on the threefold (measured by
the $r_a$'s) may lie in the marginal hyperplane and thus remain finite
in the infrared.  Thus, although the region near the original
nonsupersymmetric singular point generally inflates to infinite size,
residual supersymmetric singularities may remain at finite distance
from each other in the infrared.  Those finite distances correspond to
parameters in the infrared theory -- parameters which are clearly
determined by the corresponding marginal operators at high
energy.\footnote{Note, however, that not all high energy marginal
operators have such a geometric interpretation, and it is not clear
that the non-geometric ones survive to the infrared theory. We thank
Emil Martinec and Greg Moore for discussions on this point.}

\section{Tachyon decay products in two dimensions}

To avoid giving the impression that the phenomenon of stable and 
unstable phases, with the stable ones related by marginal operators
in the infrared, in purely a three-dimensional phenomenon, we briefly
give an example in two dimensions which exhibits the same phenomenon.

\subsection{An example}

We begin with a charge matrix
\be \label{Qia2dim}
Q_i^a = \left( \bA{ccccc} 
 1 & -1 &  6 & -6 &  0 \\ 
 0 &  0 & -1 &  2 & -1 \\
 0 & -1 &  3 & -1 &  0 \\
\eA \right) .
\ee
To clarify the geometry, we can use some (non-invertible) row operations
to find
\be
\left( \bA{ccc}
 1 &  0 & -6 \\
 3 &  0 & -6 \\
 5 & 12 & -6 \\
\eA \right)
\left( \bA{ccccc} 
 1 & -1 &  6 & -6 &  0 \\ 
 0 &  0 & -1 &  2 & -1 \\
 0 & -1 &  3 & -1 &  0 \\
\eA \right)
=
\left( \bA{ccccc} 
 1 & 5 & -12 & 0 & 0 \\
 3 & 3 & 0 & -12 & 0 \\
 5 & 1 & 0 & 0 & -12 \\
\eA \right)
\ee
This is in fact the orbifold $\BC^2/\BZ_{12}\ (1, 5)$, which has three
tachyons (visible in the second form of the charge matrix).

The classical vacua of the theory are found by studying the D-terms 
(corresponding to the first form of the charge matrix):
\bea\label{2dDterms}
D_1 &=& |\phi_1|^2 -  |\phi_2|^2 + 6 |T_1|^2 - 6 |T_2|^2 - r_1 = 0 
,
\nonumber\\
D_2 &=& - |T_1|^2 + 2 |T_2|^2 - |T_3|^2  - r_2 = 0 
,
\nonumber\\
D_3 &=& - |\phi_2|^2 + 3 |T_1|^2 - |T_2|^2  - r_3 = 0 
.
\eea
As in the previous example, the vectors bounding the phases are
$\phi_1\equiv (1,0,0)$, $\phi_2\equiv (-1,0,-1)$, $T_1\equiv (6,-1,3)$,
$T_2\equiv (-1,2,-1)$, and $T_3\equiv (0,-1,0)$ together with
$\phi_0\equiv (3,-1,0)$ which arises as the intersection of the
planes $\{\phi_1,T_3\}$ and $\{\phi_2,T_1\}$.

There are eight phases in this example: six unstable and two stable
phases.  One stable phase corresponds to a complete resolution of the
singularity, which consists of a chain of three rational curves, with
the outer curves having self-intersection $-3$ and the middle curve
having self-intersection $-2$. This follows from the Hirzebruch-Jung 
continued fraction expansion 
$${12\over 5} = 3 - \frac1{2 - \displaystyle\frac13}$$ whose coefficients 
give the self-intersections. In the other stable phase, the middle
curve is not resolved, but rather appears as a $\BC^2/\BZ_2$
singularity.  The size of the middle curve serves as the marginal
parameter in the infrared. (The two outer curves blow up to infinite
size, but the middle curve remains finite size, even in the ``smooth''
stable phase.)  In the other stable phase, the marginal parameter is a
twist field for the orbifold.

We summarize the data for this example in Table 3 similar to those of the
previous section. There are six possible types of trajectories, passing
through the various phases:
\begin{eqnarray} \nonumber
 1 \to 2 \to 7 \\ \nonumber
1 \to 4 \to 7 \\ \nonumber
1\to 4\to 6\to 8\\ 
1\to3\to 6\to8\\ \nonumber
1\to3\to5\to8\\ \nonumber
1\to2\to5\to8
\end{eqnarray}
\begin{table}[htb]
\bc
\caption{Phases of $\BC^2/\BZ_{12}\ (1,5)$}
\label{table2d}
\begin{tabular}{|c| p{5.5cm} | p{4.5cm} | p{4cm} |}
\hline
1 & $\{ T_1, T_2, T_3 \} $
& $(\phi_1,\phi_2)$
& Orbifold \\ \hline
2 & $\{ \phi_2, T_2, T_3 \} 
= \{ \phi_2, T_2, T_3 \} \cap \{ \phi_1, T_2, T_3 \} $
& $(\phi_1, T_1), \ (\phi_2, T_1)$
& Partial resolution by $T_1$ \\
\hline
3 & $\{ \phi_0, T_1, T_3 \}
= \{ \phi_1, T_1, T_3 \} \cap \{ \phi_2, T_1, T_3 \}$ 
& $(\phi_1, T_2), \ (\phi_2, T_2)$
& Partial 
resolution by $T_2$ \\
\hline
4 & $\{ \phi_1, T_1, T_2 \} 
= \{ \phi_1, T_1, T_2 \} \cap \{ \phi_2, T_1, T_2 \}$
& $(\phi_1, T_3), \ (\phi_2, T_3)$
& Partial resolution by $T_3$\\
\hline
5 & $\{ \phi_0, \phi_2, T_3 \} 
= \{\phi_1, \phi_2, T_3 \} \cap \{ \phi_1, T_2, T_3 \} \cap 
\{ \phi_2, T_1, T_3 \}$
& $(\phi_1, T_1), \ (T_1,T_2), \ (\phi_2, T_2)$
& Partial resolution by $T_1$ and $T_2$ \\ \hline
6 & $\{ \phi_0, \phi_1, T_1 \} 
= \{ \phi_1, \phi_2, T_1 \} \cap \{ \phi_2, T_1, T_2 \} \cap
\{ \phi_1, T_1, T_3 \}$
& $(\phi_1, T_3), \ (T_2, T_3), \ (\phi_2, T_2)$
& Partial resolution by $T_2$ and $T_3$ \\ \hline
7 & $\{ \phi_ 1, \phi_2, T_2 \} 
= \{ \phi_ 1, \phi_2, T_2 \} \cap \{ \phi_1, T_2, T_3 \} \cap
\{ \phi_2, T_1, T_2 \}$
& $(\phi_1, T_3), \ (T_1, T_3), \ (\phi_2,T_1)$
& Partial resolution by $T_1$ and $T_3$ \\ \hline
8 & $\{ \phi_0, \phi_1, \phi_2 \} 
= \{ \phi_1, T_2, T_3 \} \cap \{ \phi_2, T_1, T_2 \} \cap
\{ \phi_1, \phi_2, T_1 \} \cap \{ \phi_1, \phi_2, T_3 \}$
& $(\phi_1, T_3)$,  $(T_2,T_3)$, $(T_1, T_2)$,  $(\phi_2, T_1)$
& Complete resolution \\ \hline
\end{tabular}
\ec
\end{table}

\section{Conclusions and discussion}

We have seen that the geometric (Higgs branch) phases of a GLSM for a
nonsupersymmetric orbifold correspond to the possible spacetime
geometries that can result as the endpoint of tachyon condensation.
The stability of these phases is related to the possible existence of
flip transition trajectories that flow towards the more stable phases
which correspond to the condensation of the most relevant tachyon
sequence. In addition to the Higgs branch vacua, the GLSM also
exhibits nonclassical Coulomb branch vacua: it would be interesting to
interpret these from the point of view of the nonlinear sigma model
describing string propagation on the target space geometry.

\begin{figure}
\bc
\epsfig{file=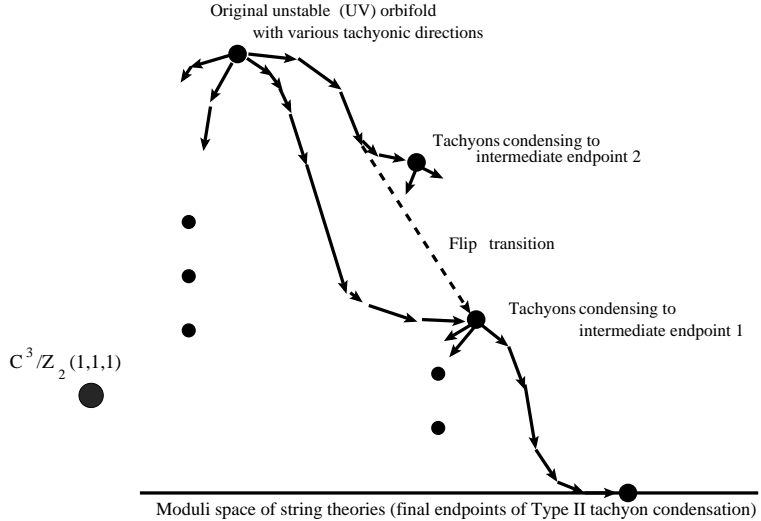, width=10cm}
\caption{A heuristic picture of an unstable (UV) orbifold with several
relevant directions: two distinct tachyons in $\BC^3/\BZ_N$ condense
to distinct endpoints, with a possible flip transition between
them. Intermediate endpoint I stemming from condensation of a more
relevant tachyon is less singular than intermediate endpoint II.
The $final$ endpoint of condensation of all tachyons for Type II
string theories is the smooth moduli space. Type 0 theories however 
admit a truly terminal singularity $\BC^3/\BZ_2\ (1,1,1)$ in the 
spectrum of decay endpoints.}
\label{figTflows}
\ec
\end{figure}
We have described flip transitions in this paper from the point of
view of the GLSM and its interrelations with the toric description in
\cite{drmknmrp}. While at the level of the GLSM, flips appear
structurally similar to flop transitions in supersymmetric
Calabi-Yaus, the physics they describe is quite different, as we have
seen. It would be interesting to understand the role of
nonperturbative (in the string coupling) effects such as D-branes in
these theories, in part with a view to understanding similarities and
differences in the physics of (nonsupersymmetric) flips and
(supersymmetric) flops. This is perhaps closely tied to a better
understanding of the geometry in the vicinity of the nonsupersymmetric
analogue of the conifold singularity itself, as well as its
deformations. It would also be interesting to obtain a better
understanding of the connection between worldsheet RG flow and direct
time evolution in spacetime, and the time evolution of flip transition
trajectories.

The GLSM analysis here gives some insight into the space of geometries
and the off-shell closed string tachyon potential. Intuitively we
expect that the distinct phases of GLSMs correspond to the various
extrema of the tachyon potential. Then thinking of an unstable
$\BC^3/\BZ_N$ orbifold as the UV fixed point with several unstable
directions (see figure~\ref{figTflows}), we know from \cite{drmknmrp}
that the $final$ endpoints of tachyon condensation do not include
terminal singularities and are always supersymmetric spaces (generally
smooth spaces such as supersymmetric Calabi-Yau manifolds, with
supersymmetric singularities also permitted) for Type II string
theories. These final endpoints can be thought of as the culmination
of condensation of all tachyons, chiral and nonchiral\footnote{As seen
in \cite{drmknmrp} (in the Type II example discussed there),
condensation of purely chiral tachyons will in general result in
geometric terminal singularities which will then continue to decay via
the nonchiral tachyonic blowup modes to finally result in
supersymmetric spaces.}, and are the absolute minima of the tachyon
potential. Thus the depth of the tachyon potential for a given
orbifold is basically the height of the orbifold ``hill'' maximum
above the absolute minima, \ie, the (smooth) moduli space.

In an off-shell formulation of string theory, one expects that the 
tachyon potential\footnote{Note that the status of the tachyon 
potential for noncompact codimension three and two singularities is 
somewhat different from that for $\BC/\BZ_N$: under tachyon 
condensation, while the asymptotic geometry remains an $S^1$ for the 
latter, the asymptotics changes for $\BC^3/\BZ_N$ and $\BC^2/\BZ_N$, 
making it hard to compare, say, energies of the geometries at early and 
late times. The heuristic picture of the tachyon potential in 
figure~\ref{figTflows} is thus best interpreted as a formal picture of 
the ``space of string theories'', without a precise quantitative 
physical measure thereof, at least for the present.} for a given 
orbifold $\BC^3/\BZ_N\ (1,p,q)$ is a function $V(T)$ describing the 
dynamics of all the tachyon fields $T_i$ in the system, with parameters 
$N, p, q$. This is to be treated with some care since a given $\BZ_N\ 
(1,p,q)$ orbifold can be relabelled in terms of various distinct 
parameters $p,q$, so that naive expressions for $V(T)$ might in fact be 
incorrect. \cite{atish0111} gives a conjecture for the depth of the 
potential $V(T)$ for $\BC/\BZ_N$. However unlike $\BC/\BZ_N$, where the 
most relevant tachyon decay channel leads to flat space, $\BC^2/\BZ_N$ 
and $\BC^3/\BZ_N$ orbifolds typically do not decay to supersymmetric
spaces via the most relevant tachyon: the endpoints themselves are
typically unstable to tachyon condensation via subsequent tachyons.
Expressions involving condensation of a single tachyon appear in
\cite{atishvafa0111, ssin0308015, sarkar0407}. Using these, one can
thus attempt an iterative guess for the depth $V(T)$ by summing over
all tachyons (mass $M^2$, R-charge $R$) present in the orbifold,
\be
V(T) \sim \sum_k M^2_{T_k} \sim \sum_k (R_k - 1)\ .
\ee
It is important to realize that subsequent tachyons can potentially 
become irrelevant in $\BC^3/\BZ_N$ \cite{drmknmrp}, so that care must 
be used in evaluating the sum in this expression. It is then reasonable 
to expect that this sum in fact converges to a quantity that involves 
only the parameters $N, p, q$ for the original unstable orbifold that 
we began with (note that this final expression must be independent of 
possible redefinitions of the $k_i$ labelling the same physical orbifold): 
it is further tempting to ask if there are connections to the $g_{cl}$
conjecture of \cite{hkmm}. It would be interesting to pursue these
lines of thought further to get deeper insights into the tachyon
potential and the ``space of string theories''.

\vspace{15mm}

{\small {\bf Acknowledgments:} It is a pleasure to thank Allan Adams, Paul 
Aspinwall, Atish Dabholkar, Emil Martinec, Ilarion Melnikov, Shiraz 
Minwalla, Greg Moore, Sunil Mukhi and Ronen Plesser for useful discussions. 
This work is partially supported by NSF grant DMS-0301476.}

\end{document}